\documentclass[structabstract]{aa}  
\usepackage{graphicx}
\usepackage{lscape}
\usepackage{subfig}
\usepackage{float}
\usepackage{natbib}
\usepackage{txfonts}
\begin{document}
   \title{Secular- and merger-built bulges in barred galaxies}

   \author{J. M\'endez-Abreu
          \inst{1,2,3}\fnmsep\thanks{Juan de la Cierva Fellow}
          \and
          Victor P. Debattista
          \inst{4}
          \and
          E. M. Corsini
          \inst{5,6}
          \and
          J. A. L. Aguerri
          \inst{1,2}}

   \institute{Instituto de Astrof\'isica de Canarias, Calle V\'ia L\'actea s/n, E-38200 La Laguna, Tenerife, Spain
         \and
            Departamento de Astrof\'isica, Universidad de La Laguna,  E-38205 La Laguna, Tenerife, Spain
            \and
            School of Physics and Astronomy, University of St Andrews, SUPA, North Haugh, KY16 9SS St Andrews, UK \\
 \email{jma20@st-andrews.ac.uk}
            \and
            Jeremiah Horrocks Institute, University of Central Lancashire, PR1 2HE Preston, UK
            \and
            Dipartimento di Fisica e Astronomia `G. Galilei', Universit\`a di Padova, vicolo dell'Osservatorio 3, I-35122 Padova,
Italy
            \and
            INAF - Osservatorio Astronomico di Padova, vicolo dell'Osservatorio 5, I-35122 Padova, Italy
             }

   \date{Received September 15, 1996; accepted March 16, 1997}

 
  \abstract
%
{Historically, galaxy bulges were  thought of single-component objects
  at the center of galaxies.   However, this picture is now questioned
  since different bulge types, namely classical and pseudobulges, with
  different formation paths have been found coexisting within the same
  galaxy.}
%
%
{We study the incidence, as well as the nature, of composite bulges in
  a sample  of 10 face-on  barred galaxies to constrain  the formation
  and evolutionary processes of the central regions of disk galaxies.}
%
%
{We analyze  the morphological, photometric, and  kinematic properties
  of each  bulge. Then, by  using a case-by-case analysis  we identify
  composite bulges  and classify every  component into a  classical or
  pseudobulge.  In addition,  bar-related boxy/peanut (B/P) structures
  were also identified and characterized.}
%
%
{We find  only three  galaxies hosting  a single-component  bulge (two
  pseudobulges  and one  classical bulge).   Thus, we  demonstrate the
  high incidence of composite bulges (70\%) in barred galaxies.
We find evidence of composite bulges coming in two main types based on
their formation: secular-built and merger- and secular-built.
We  call  secular-built  to  composite  bulges  made  of  entirely  by
structures  associated with  secular processes  such as  pseudobulges,
central disks, or  B/P bulges.  We find four composite  bulges of this
kind in our sample.
On the  other hand, merger-  and secular-built bulges are  those where
structures  with different  formation  paths coexist  within the  same
galaxy,  i.e.,  a  classical  bulge coexisting  with  a  secular-built
structure (pseudobulge, central  disk, or B/P).  Three  bulges of this
kind were found in the sample.
We remark on the importance  of detecting kinematic structures such as
$\sigma-$drops  to  identify  composite   bulges.   A  large  fraction
($\sim$80\%)  of  galaxies  were   found  to  host  $\sigma-$drops  or
$\sigma-$plateaus  in our  sample  revealing their  high incidence  in
barred galaxies.  }
%
%
{The  high frequency  of composite  bulges in  barred galaxies  points
  towards a complex formation and evolutionary scenario. Moreover, the
  evidence for  coexisting merger- and secular-built  bulges reinforce
  this idea.   We discuss how  the presence of different  bulge types,
  with  different formation  histories and  timescales, can  constrain
  current models of bulge formation.

}

   \maketitle  

\section{Introduction}
\label{sec:intro}

The central role  of the bulges of disk galaxies  on our understanding
of galaxy  formation and evolution  is a generally accepted  fact. The
bulge prominence is the main  feature by which galaxies are classified
into Hubble types, accounting for more than one quarter of the stellar
mass budget in the  local Universe \citep{driver07,gadotti09}.  Bulges
influence    the   size,    strength,    and    incidence   of    bars
\citep{sellwood81,laurikainen09,aguerri09},  which  in   turn  have  a
profound  effect  on  the  rest   of  the  stellar  and  gaseous  mass
distribution.   Bulges host  central supermassive  black holes,  which
themselves   influence    galaxy   evolution   and    star   formation
\citep{springel05}.  Remarkably,  the mass  of the  supermassive black
hole  correlates  with  the  random  motions of  stars  in  the  bulge
\citep{gebhardt00}, suggesting that these  objects of vastly different
scales somehow regulate each other \citep[but see][]{kormendy11}.
However, despite their importance, questions such as how much of their
observed morphology is due to  fast/long-term processes or how much is
due to internal/external evolution  remain unanswered.  In fact, these
different formation processes are reflected in the two types of bulges
which are thought to exists: classical and disk-like bulges.

Classical   bulges  can   be  formed   via  dissipative   collapse  of
protogalactic  gas clouds  \citep{eggen62}  or by  the coalescence  of
giant  clumps in  primordial disks  \citep{bournaud07}.  On  the other
hand, they could  also grow out of disk  material externally triggered
by    satellite     accretion    during    minor     merging    events
\citep{aguerri01,elichemoral06}      or     by      galaxy     mergers
\citep{kauffmann96} with different merger histories \citep{hopkins10}.
Extreme cases  of polar bulges sticking  out from the disk  plane have
also  been   observed  \citep[][and   references  therein]{corsini12}.
Observationally,  the  surface-brightness  distribution  of  classical
bulges generally  follows a  de Vaucouleurs  law \citep{andredakis95}.
They appear rounder  than their associated disks  and their kinematics
are  well described  by rotationally  flattened oblate  spheroids with
little or  no anisotropy \citep{kormendyillingworth82}.   In addition,
they  have  photometric and  kinematic  properties  which satisfy  the
fundamental             plane             (FP)             correlation
\citep{bender92,falconbarroso02,aguerri05}.

Disk-like bulges  (also known as  pseudobulges) are thought to  be the
products     of      secular     processes     driven      by     bars
\citep[see][]{kormendykennicutt04}.   Bars  are   ubiquitous  in  disk
galaxies \citep{aguerri09}. They are  efficient mechanisms for driving
gas inward  to the galactic  center triggering central  star formation
generally    associated    with    a    pseudobulge.     Nevertheless,
\citet{elichemoral11} recently  proposed that pseudobulges  might also
be created by the secular accretion of low-density satellites into the
main galaxy,  thus providing  an alternative  to bar-driven  growth of
pseudobulges. Observationally, pseudobulges have an almost exponential
surface-brightness  distribution \citep{fisherdrory08},  and they  are
mainly   oblate   spheroids    \citep{mendezabreu10}   with   apparent
flattenings similar to their associated disks.  Pseudobulges rotate as
fast as disks and usually deviate from the FP \citep{carollo99}.

Other  central  structures  usually  referred  to  as  bulges  include
boxy/peanut (B/P) bulges.  It is  nowadays well established that these
are   structures   related   to   the  secular   evolution   of   bars
\citep{combessanders81,chungbureau04}.  As  bars evolve, stars  can be
moved perpendicular to the disk plane due to a coherent bending of the
bar         producing         its         characteristic         shape
\citep{debattista04,martinezvalpuesta06}.   Therefore, B/P  structures
share   the    photometric   and   kinematic   properties    of   bars
\citep{mendezabreu08b,erwindebattista13}.

Mixed types of  classical and pseudobulges can also occur  in the same
galaxy   \citep{athanassoula05,gadotti09,nowak10,kormendybarentine10}.
Furthermore, their combination with  B/P structures in barred galaxies
is also  expected.  The frequency  of these composite bulges  is still
not well-determined,  and, more puzzlingly, different  combinations of
bulge     types      can     successfully      explain     observations
\citep{delorenzocaceres12}.  Besides  bulges, a variety  of structures
present  in the  center  of  disk galaxies  have  been found:  nuclear
stellar  disks,  characterized by  a  smaller  scale-length and  higher
central surface-brightness with respect  to the large kiloparsec scale
disks  typical  of  lenticular  and  spiral  galaxies  \citep[see][and
  references   therein]{morelli10};  nuclear   clusters,  are   common
structures in late Hubble types with linear scales from a few up to 20
pc \citep{boker02,seth06}; and nuclear  bars, subkiloparsec scale bars
usually     lying    inside     typical     large    galactic     bars
\citep{prieto97,laine02,erwin04}.  This complex picture of the central
regions of disk galaxies,  containing several structural and kinematic
components   formed  through   very   different  mechanisms,   further
complicate the identification of different bulge types.

One  promising path  to  uncover the  real nature  of  bulges in  disk
galaxies  is  through  the  combination of  structural  and  kinematic
diagnostics.   Thus   far,  the  most  commonly   used  diagnostic  to
distinguish bulge  types was  based only  on the  photometry, assuming
that  light profiles  of classical  bulges are  better described  by a
S\'ersic profile with  $n > 2$, whereas pseudobulges should  have $n <
2$ \citep{fisherdrory08}.   Recently, \citet{fabricius12}  studied the
photometric and kinematic properties for a sample of 45 bulges in disk
galaxies.  They demonstrate that  these combined properties are useful
to distinguish classical from pseudobulges  even if some unclear cases
are still present.

In  this  paper  we  present  intermediate-resolution  kinematics  and
structural  properties of  10 face-on  barred galaxies.   We use  this
information to reveal  the nature of the  different components present
in  the galaxy  centers, and  more  specifically the  nature of  their
bulges.  Kinematic measurements of face-on galaxies, and in particular
of   barred   galaxies  are   rare   in   the  literature   \citep[but
  see][]{bershady10}.  They are usually  avoided since problems due to
the  effects  of  projection  on  the  velocity  rotation  curves  are
difficult  to overcome.   However, stellar  kinematic measurements  of
face-on galaxies provide access to  the vertical component of both the
velocity  ellipsoid and  higher order  moments  of the  line of  sight
velocity   distribution   (LOSVD)   providing   new   hints   to   the
identification and understanding of galaxy structural components.

The paper is organized as follow: Sect.~\ref{sec:sample} describes the
sample        selection.        Sects.~\ref{sec:photometry}        and
\ref{sec:spectroscopy}  detail the  observations,  data reduction  and
analysis  of the  photometric  and  spectroscopic data,  respectively.
Sect.~\ref{sec:results} presents  the analysis  of the  structural and
kinematic  properties   of  the   bulges.   Sect.~\ref{sec:discussion}
describes  the  different bulge  types  present  in our  sample.   The
conclusions      are     given      in     Sect.~\ref{sec:conclusion}.
Appendix~\ref{sec:notes} summarizes  the main characteristics  of each
sample galaxy. Throughout the article  we assume a flat cosmology with
$\Omega_{\rm m}$  = 0.3,  $\Omega_{\rm \Lambda}$ =  0.7, and  a Hubble
constant $H_0$ = 70 km s$^{-1}$ Mpc$^{-1}$.

\section{Sample selection}
\label{sec:sample}

The sample of face-on barred  galaxies was selected from the NASA/IPAC
Extragalactic Database  (NED) as bright  ($B_T < 14$)  and undisturbed
objects, with a  disk inclination lower than $30^\circ$,  a bar length
larger  than $10\arcsec$  to  avoid seeing  effects,  a disk  diameter
smaller  than  $4\arcmin$  to  allow  an  accurate  spectroscopic  sky
subtraction, no strong evidence of dust and no bright foreground stars
in the Digitized  Sky Survey (DSS) image.  To define  the final galaxy
sample, their disk inclinations, bar  lengths, and bar position angles
were first  determined from  ellipse fits to  the Two-Microns  All Sky
Survey (2MASS) Large Galaxy  Atlas \citep{jarrett03} $J-$band archival
images following the method  described in \citet{aguerri09}.  The main
properties of the sample galaxies are given in Table \ref{tab:sample}.

\begin{table}
\caption{Characteristics of the sample galaxies}   
\label{tab:sample}    
\centering            
\begin{tabular}{c c c c c c}      
\hline\hline     
Galaxy      &     RA      & DEC        & Morph. Type  & $z$ & $i$\\
            &   (h:m:s)   & ($^\circ$:':'') &           &     & ($^\circ$) \\
 (1)        &    (2)      &  (3)       & (4) &  (5)     &  (6)\\
\hline
   IC~1815  & 02:34:20.0 & +32:25:46  &  SB0         & 0.0157 & 30\\
   NGC~0043 & 00:13:00.7 & +30:54:55  &  SB0         & 0.0160 & 25\\     
   NGC~0098 & 00:22:49.5 & -45:16:08  &  SB(r)bc     & 0.0206 & 22\\     
   NGC~0175 & 00:37:21.5 & -19:56:03  &  SB(r)ab     & 0.0130 & 20\\
   NGC~0521 & 01:24:33.8 & +01:43:53  &  SB(r)bc     & 0.0167 & 14\\
   NGC~0621 & 01:36:49.0 & +35:30:44  &  SB0         & 0.0172 & 34\\  
   NGC~1640 & 04:42:14.5 & -20:26:05  &  SB(r)b      & 0.0054 & 20\\
   NGC~2493 & 08:00:23.6 & +39:49:50  &  SB0         & 0.0130 & 20\\
   NGC~4477 & 12:30:02.2 & +13:38:12  &  SB(s)0:?    & 0.0045 & 27\\
   NGC~4838 & 12:57:56.1 & -13:03:36  &  (R')SB(r)b: & 0.0167 & 22\\
\hline                                 
\end{tabular}
\tablefoot{(1)  Galaxy   name;  (2)   and  (3)  right   ascension  and
  declination   of   the   galaxies   (J2000.0);   (4)   morphological
  classification; (5)  redshift; (6) inclination derived  from $q_{\rm
    disk}$ in  Table~\ref{tab:decomp}.  (2-5)  values were  taken from
  the NASA/IPAC Extragalactic Database (NED).}
\end{table}
  
\section{Surface photometry}
\label{sec:photometry}
\subsection{Observations and data reduction}

Imaging  of the  sample  galaxies was  obtained  from three  different
sources (see Table~\ref{tab:phot}): the  Sloan Digital Sky Survey Data
Release 8 \citep[SDSS-DR8;][]{aihara11},  the acquisition images taken
at  the time  of  the  spectroscopic observations  at  the Very  Large
Telescope (VLT), and 2MASS \citep[][]{skrutskie06}.

For the purpose of this paper, we preferred to use the images from the
SDSS since they are well calibrated and with sufficient resolution and
depth     to     safely     obtain     an     accurate     photometric
decomposition. Unfortunately, they were  only available for 5 galaxies
in our sample.  In our analysis we used the $i-$band images.

When the SDSS images were  not available, we analyzed the uncalibrated
acquisition  images  from  the   VLT.   The  acquisition  images  were
deliberately  taken  using the  $I-$band  Bessel  filter and  with  an
exposure time  usually of a few  minutes.  We reduced this  data using
the bias  and dome flat-field  images available for  the corresponding
nights, however, the photometric calibration  was not performed due to
the  lack of  photometric standards  stars.  It  is worth  noting that
absolute calibration is not necessary  to obtain the galaxy structural
parameters used in this work.  Three galaxies (NGC~0098, NGC~1640, and
NGC~4838) of the sample were analyzed using these images.

For  the remaining  two  galaxies  (IC~1815 and  NGC~0621)  we use  the
$K-$band images available  in the 2MASS survey. Even  though the image
quality and depth of these images  is clearly poorer than the previous
ones, in \citet{mendezabreu08a} we already demonstrated the usefulness
of  2MASS imaging  for photometric  decompositions of  bright galaxies
such as those  in this work. Therefore, for the  sake of completeness,
and after a  careful visual inspection of the fits  and the residuals,
we included the corresponding analysis in the present study.

\begin{table}
\caption{Characteristics of the photometric observations}   
\label{tab:phot}    
\centering            
\begin{tabular}{c c c c}      
\hline\hline              
Galaxy      &  Filter     & Source & FWHM \\
            &             &        & (\arcsec)\\   
   (1)      &    (2)      &    (3)    &  (4) \\
\hline
   IC~1815  & $K$         & 2MASS     & 3.1 \\
   NGC~0043 & $i$         & SDSS      & 1.2 \\
   NGC~0098 & $I-Bessel$  & VLT       & 0.7 \\
   NGC~0175 & $i$         & SDSS      & 1.2 \\
   NGC~0521 & $i$         & SDSS      & 0.8 \\
   NGC~0621 & $K$         & 2MASS     & 2.9 \\
   NGC~1640 & $I-Bessel$  & VLT       & 0.7 \\
   NGC~2493 & $i$         & SDSS      & 1.1 \\
   NGC~4477 & $i$         & SDSS      & 1.2 \\
   NGC~4838 & $I-Bessel$  & VLT       & 0.8 \\
\hline                                 
\end{tabular}
\tablefoot{(1) Galaxy  name; (2)  image pass-band;  (3) source  of the
  image; (4) PSF FWHM measured from field stars on the galaxy images.}
\end{table}
  
\subsection{Photometric decomposition}
\label{sec:decomp}

The  structural parameters  of  the sample  galaxies  were derived  by
applying  a two-dimensional  photometric decomposition  to the  galaxy
images previously  described. Several  codes are available  to perform
photometric  decompositions  such   as  GALFIT  \citep{peng02},  GIM2D
\citep{simard02},  or BUDDA  \citep{desouza04}.   We  used the  GASP2D
algorithm  developed  by   \citet{mendezabreu08a}.   We  will  briefly
describe in  the following  the main  characteristics of  GASP2D.  The
galaxy surface-brightness distribution (SBD) was assumed to be the sum
of a bulge,  disk, and bar component.  No  other additional components
such  as  lenses or  ovals  present  in  a  fraction of  bar  galaxies
\citep{prieto01,laurikainen09}  were  considered  in  our  photometric
decomposition.

The S\'ersic law \citep{sersic68}, also  known as the $r^{1/n}$ law or
generalized de  Vaucouleurs law, was  adopted to describe  the surface
brightness of the bulge component

\begin{equation} 
I_{\rm bulge}(r_{\rm bulge})=I_{\rm e}10^{-b_n\left[\left(\frac{r_{\rm bulge}}{r_{\rm e}} 
\right)^{\frac{1}{n}}-1\right]}, 
\label{eqn:bulge_surfbright} 
\end{equation} 
%
where  $r_{\rm  bulge}$  is  the  radius  measured  in  the  Cartesian
coordinates describing the reference system  of the bulge in the plane
of the  sky. $r_{\rm e}$, $I_{\rm  e}$, and $n$ are  the effective (or
half-light) radius, the surface brightness at $r_{\rm e}$, and a shape
parameter describing the curvature  of the SBD, respectively, and $b_n
\simeq 0.868\,n-0.142$ \citep{caon93}.  The bulge isophotes are ellipses
centered on  the galaxy center ($x_0$, $y_0$),  with constant position
angle $PA_{\rm bulge}$ and constant axial ratio $q_{\rm bulge}$.

The SBD of the disk component was assumed to follow an exponential law
\citep{freeman70}

\begin{equation} 
I_{\rm disk}(r_{\rm disk}) = I_0\,e^{-\left(\frac{r_{\rm disk}}{h}\right)}, 
\label{eqn:disc_surfbright} 
\end{equation} 
%
where  $r_{\rm  disk}$  is   the  radius  measured  in  the  Cartesian
coordinates describing the reference system of the disk. $I_0$ and $h$
are  the central  surface  brightness and  scale-length  of the  disk,
respectively.   The disk  isophotes are  ellipses centered  on ($x_0$,
$y_0$),  with constant  position  angle $PA_{\rm  disk}$ and  constant
axial ratio $q_{\rm disk}$.

The projected surface density of a three-dimensional Ferrers ellipsoid
(\citealt{ferrers77},  see  also  \citealt{aguerri09}) was  used  to
describe the SBD of the bar component

\begin{equation}
I_{\rm bar}(r_{\rm bar})=I_{\rm 0,bar}\left[1-\left(\frac{r_{\rm bar}}{a_{\rm bar}}\right)^2\right]^{n_{\rm bar}+0.5}; 
\qquad r_{\rm bar} \le a_{\rm bar},
\end{equation}
%
where  $r_{\rm  bar  }$  is  the  radius  measured  in  the  Cartesian
coordinates  describing the  reference  system of  the  bar and  using
generalized ellipses  \citep{athanassoula90}. $I_{\rm 0,bar}$, $a_{\rm
  bar}$ and  $n_{\rm bar}$  represent the central  surface brightness,
length and shape  parameter of the bar, respectively.  Due to the high
degree  of  degeneracy that  the  $n_{\rm  bar}$ parameter  introduces
during the fit, we decided to  keep it as a fixed parameter during the
fitting process. The default value used was $n_{\rm bar}=2$ \citep[see
  also][]{laurikainen05}.

The bar strength  represents the contribution of the bar  to the total
galaxy  potential. We  compute the  bar strength  using the  recipe of
\citet{whyte02} which is based on the bar axis ratio

\begin{equation}
f_{\rm b} = \frac{2}{\pi} \left(\arctan \left[q_{\rm bar}^{-1/2}\right]
  - \arctan \left[q_{\rm bar}^{+1/2}\right] \right),
\end{equation}
%
where  $q_{bar}$ is  the  axial ratio  of the  bar  corrected for  the
inclination  of the  disk \citep[see][]{abraham99}.   We used  the bar
ellipticities measured  from our photometric decomposition,  which can
strongly differ  from those  derived from the  maximum of  the ellipse
averaged isophotal radial  profiles \citep[see][]{gadotti08}.

To derive  the photometric parameters  of the different  components we
iteratively fitted  a model  of the  SBD to the  pixels of  the galaxy
image, using a non-linear least-squares minimization based on a robust
Levenberg-Marquardt method \citep{more80}.  The actual computation has
been done using the MPFIT algorithm \citep{marquardt09} under the IDL
\footnote{Interactive  Data  Language  is distributed  by  ITT  Visual
  Information Solutions.  It  is available from http://www.ittvis.com}
environment.   Each image  pixel has  been weighted  according to  the
variance of its  total observed photon counts due  to the contribution
of  both the  galaxy and  sky,  and determined  assuming photon  noise
limitation and  taking into  account the  detector readout  noise. The
seeing effects were  taken into account by convolving  the model image
with a circular Moffat  \citep{trujillo01} point spread function (PSF)
with  the full  width at  half maximum  (FWHM) measured  directly from
stars in the galaxy image (Table~\ref{tab:phot}).

Figure~\ref{fig:phot} shows  the GASP2D fits  for each galaxy  in the
sample.   The parameters  derived  for the  structural components  are
collected in Table~\ref{tab:decomp}.

The formal  errors obtained  from the $\chi^2$  minimization procedure
are usually  not representative of  the real errors in  the structural
parameters  \citep{mendezabreu08a}.  Therefore,  the  errors given  in
Table \ref{tab:decomp} were  obtained through a series  of Monte Carlo
simulations.  A set  of 500 images of galaxies with  a S\'ersic bulge,
an exponential disk, and a  Ferrers bar was generated.  The structural
parameters of the  artificial galaxies were randomly  chosen among the
ranges obtained for our sample galaxies (Table~\ref{tab:decomp}).  The
simulated galaxies were  assumed to be at a distance  of 51 Mpc, which
corresponds  to the  mean of  our  galaxy sample.   The adopted  pixel
scale,  CCD  gain,  and  read-out-noise   were  chosen  to  mimic  the
instrumental  setup  of  the  photometric  observations.   Finally,  a
background level and photon noise  were added to the artificial images
to yield a signal-to-noise ratio similar to that of the observed ones.
The images of artificial galaxies were analyzed with GASP2D as if they
were real.  Thus, the initial conditions in the fitting procedure were
computed independent  of their  actual values  for each  model galaxy.
The errors  on the fitted  parameters were estimated by  comparing the
input  and measured  values assuming  they were  normally distributed.
The  mean  and  standard  deviation  of the  relative  errors  of  the
artificial galaxies were adopted as  the systematic and typical errors
for the observed galaxies.  It  is worth noting that systematic errors
associated with  PSF and sky  background uncertainties were  not taken
into account in the simulations  and therefore the quoted errors might
still be underestimated \citep[e.g.,][]{mendezabreu08a}.

\begin{landscape}
\begin{table}
\caption{Structural parameters of the sample galaxies}   
\label{tab:decomp}    
\begin{tabular}{c c c c c c c c c c c c c c}      
\hline\hline              
Galaxy        &   $r_{\rm e}$   & $n$  & $q_{\rm bulge}$ & $PA_{\rm bulge}$ & $h$ & $q_{\rm disk}$ & $PA_{\rm disk}$ & $a_{\rm bar}$ & $q_{\rm bar}$ & $PA_{\rm bar}$ & $B/T$ & $Bar/T$ & $r_{\rm bd}$ \\   
              &     (\arcsec)      &      &                &   ($^{\circ}$)   &(\arcsec)&                &  ($^{\circ}$)    &  (\arcsec)       &                 &  ($^{\circ}$)   & & & (\arcsec) \\
(1) & (2) & (3) & (4) & (5) & (6) & (7) & (8) & (9) & (10) & (11) & (12) & (13) & (14)\\
\hline
IC 1815   &   2.9$\pm$ 0.3  &   3.4$\pm$ 0.5  &  0.79$\pm$ 0.08  &     120.8$\pm$ 13.2  &    15.4$\pm$ 3.1  &  0.87$\pm$  0.17  &     115.8$\pm$  23   &    17.2$\pm$  2.5  &  0.32$\pm$  0.04  &     132.2$\pm$  19   &  0.33  &  0.09  &  5.00  \\
NGC 0043  &   2.3$\pm$ 0.1  &   2.2$\pm$ 0.2  &  0.87$\pm$ 0.04  &     119.0$\pm$  6.0  &    14.2$\pm$ 0.7  &  0.91$\pm$  0.04  &     157.3$\pm$  7.8  &    19.9$\pm$  1.3  &  0.44$\pm$  0.03  &      94.4$\pm$  6.6  &  0.21  &  0.07  &  5.52  \\
NGC 0098  &   1.6$\pm$ 0.1  &   1.2$\pm$ 0.1  &  0.73$\pm$ 0.04  &     206.6$\pm$ 12.3  &     9.6$\pm$ 0.6  &  0.93$\pm$  0.05  &     201.6$\pm$  12   &    26.1$\pm$  2.0  &  0.20$\pm$  0.01  &      36.5$\pm$  2.9  &  0.11  &  0.08  &  3.25  \\
NGC 0175  &   3.0$\pm$ 0.3  &   1.3$\pm$ 0.2  &  0.71$\pm$ 0.07  &     124.4$\pm$ 13.6  &    20.3$\pm$ 2.2  &  0.94$\pm$  0.09  &      66.7$\pm$  6.6  &    43.8$\pm$  6.5  &  0.20$\pm$  0.02  &     123.3$\pm$  12   &  0.07  &  0.11  &  5.12  \\
NGC 0521  &   3.6$\pm$ 0.4  &   2.4$\pm$ 0.3  &  0.94$\pm$ 0.10  &     167.4$\pm$ 18.4  &    36.6$\pm$ 4.0  &  0.97$\pm$  0.09  &      21.2$\pm$  2.1  &    28.5$\pm$  4.2  &  0.26$\pm$  0.02  &     156.9$\pm$  15   &  0.08  &  0.02  &  6.70  \\
NGC 0621  &   1.6$\pm$ 0.2  &   1.7$\pm$ 0.2  &  0.89$\pm$ 0.08  &      62.4$\pm$  6.9  &    13.3$\pm$ 2.7  &  0.83$\pm$  0.16  &      34.5$\pm$  6.8  &    19.5$\pm$  2.9  &  0.35$\pm$  0.05  &     104.4$\pm$  15   &  0.27  &  0.14  &  4.00  \\
NGC 1640  &   3.4$\pm$ 0.2  &   1.9$\pm$ 0.2  &  0.69$\pm$ 0.03  &     220.1$\pm$ 11.0  &    16.6$\pm$ 0.8  &  0.94$\pm$  0.04  &     225.1$\pm$  11   &    46.0$\pm$  2.3  &  0.19$\pm$  0.00  &     223.9$\pm$  11   &  0.09  &  0.07  &  5.00  \\
NGC 2493  &   4.6$\pm$ 0.2  &   2.0$\pm$ 0.2  &  0.88$\pm$ 0.04  &      44.6$\pm$  2.2  &    23.5$\pm$ 1.2  &  0.94$\pm$  0.04  &     100.8$\pm$  5.0  &    34.0$\pm$  1.7  &  0.36$\pm$  0.01  &      31.8$\pm$  2.0  &  0.26  &  0.10  &  8.68  \\
NGC 4477  &   6.1$\pm$ 0.2  &   1.6$\pm$ 0.1  &  0.86$\pm$ 0.02  &      24.6$\pm$  0.7  &    32.2$\pm$ 1.0  &  0.89$\pm$  0.02  &      61.1$\pm$  1.8  &    45.6$\pm$  1.5  &  0.39$\pm$  0.01  &      10.2$\pm$  0.3  &  0.20  &  0.08  &  12.2  \\
NGC 4838  &   1.8$\pm$ 0.1  &   1.4$\pm$ 0.1  &  0.83$\pm$ 0.05  &     219.0$\pm$ 13.1  &    11.2$\pm$ 0.7  &  0.93$\pm$  0.05  &     162.2$\pm$  9.7  &    25.2$\pm$  1.5  &  0.31$\pm$  0.01  &      54.8$\pm$  3.2  &  0.10  &  0.12  &  3.25  \\
\hline                                 
\end{tabular}
\tablefoot{(1) Galaxy  name; (2), (3), (4), and  (5) effective radius,
  shape  parameter,  axis ratio,  and  position  angle  of the  bulge,
  respectively;  (6),  (7), and  (8)  scale-length,  axis ratio,  and
  position  angle  of the  disk,  respectively;  (9),  (10), and  (11)
  semi-major axis length,  axis ratio, and position angle  of the bar,
  respectively;    (12)   bulge-to-total   luminosity    ratio;   (13)
  bar-to-total luminosity  ratio; (14) radius where the  bulge and the
  disk  (or bar)  component give  the same  contribution to  the total
  surface brightness.}
\end{table}
\end{landscape}

   \begin{figure*}
   \centering
   \includegraphics[angle=90,width=0.49\textwidth]{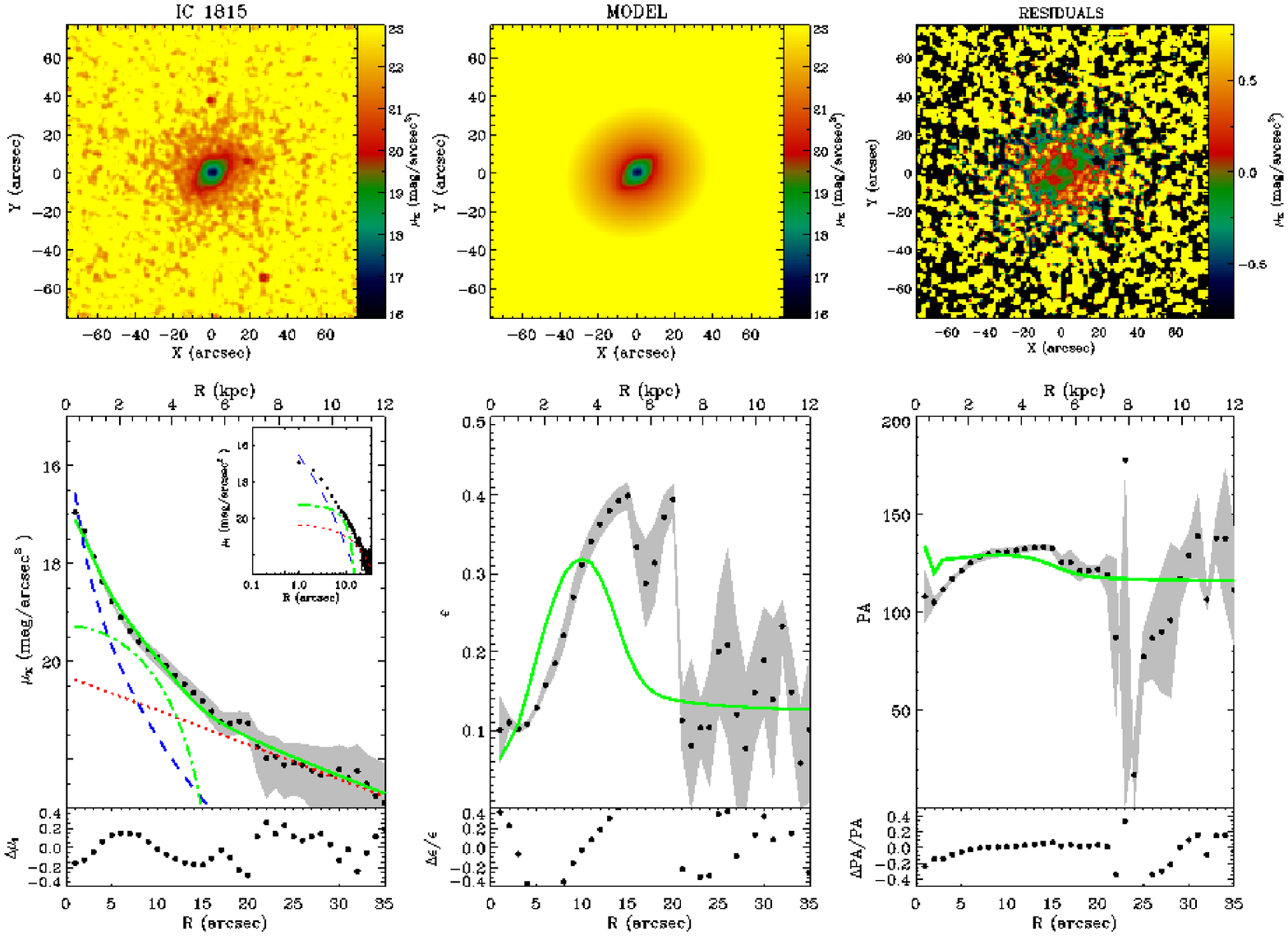}
   \includegraphics[angle=90,width=0.49\textwidth]{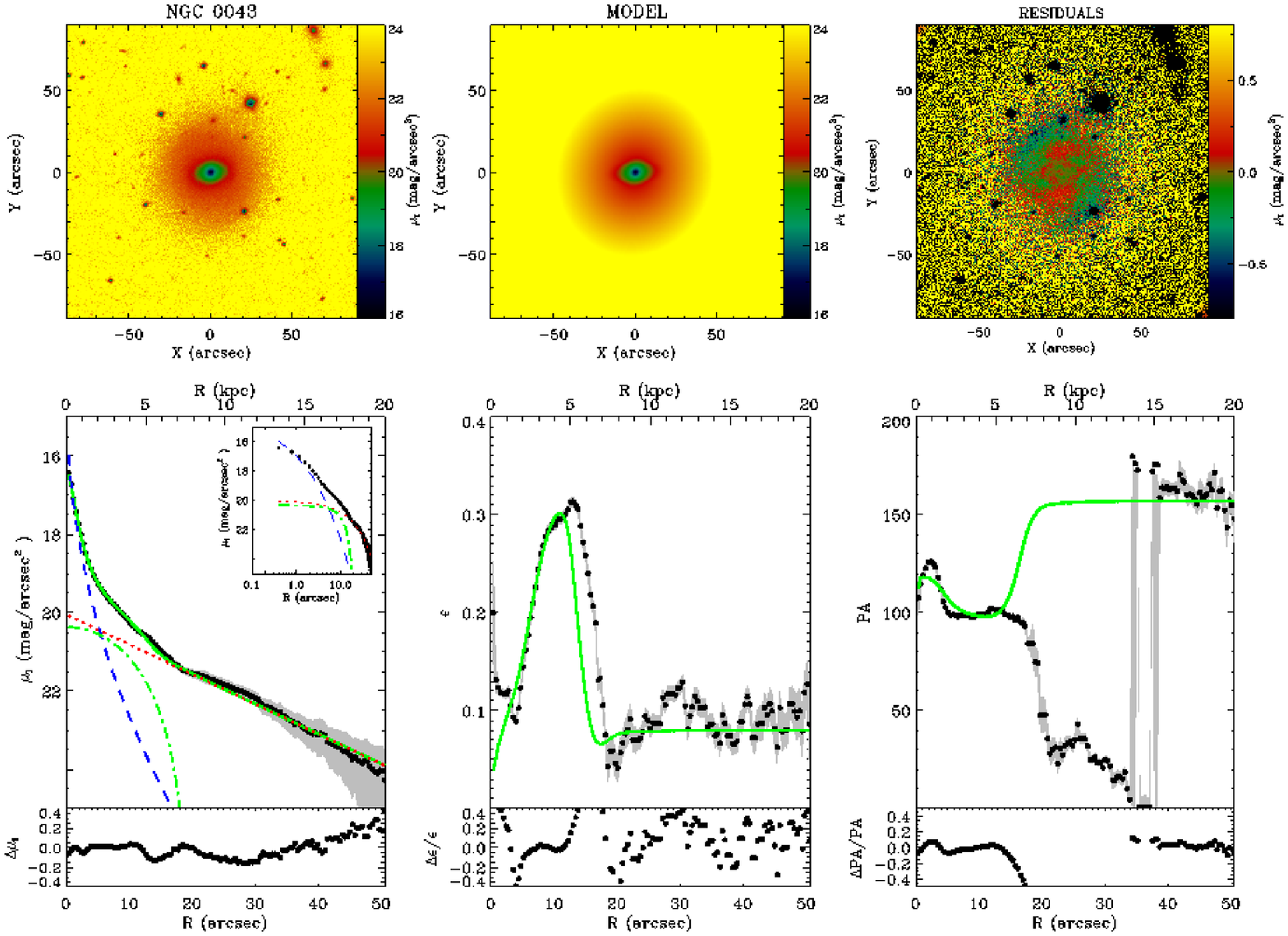}
   \includegraphics[angle=90,width=0.49\textwidth]{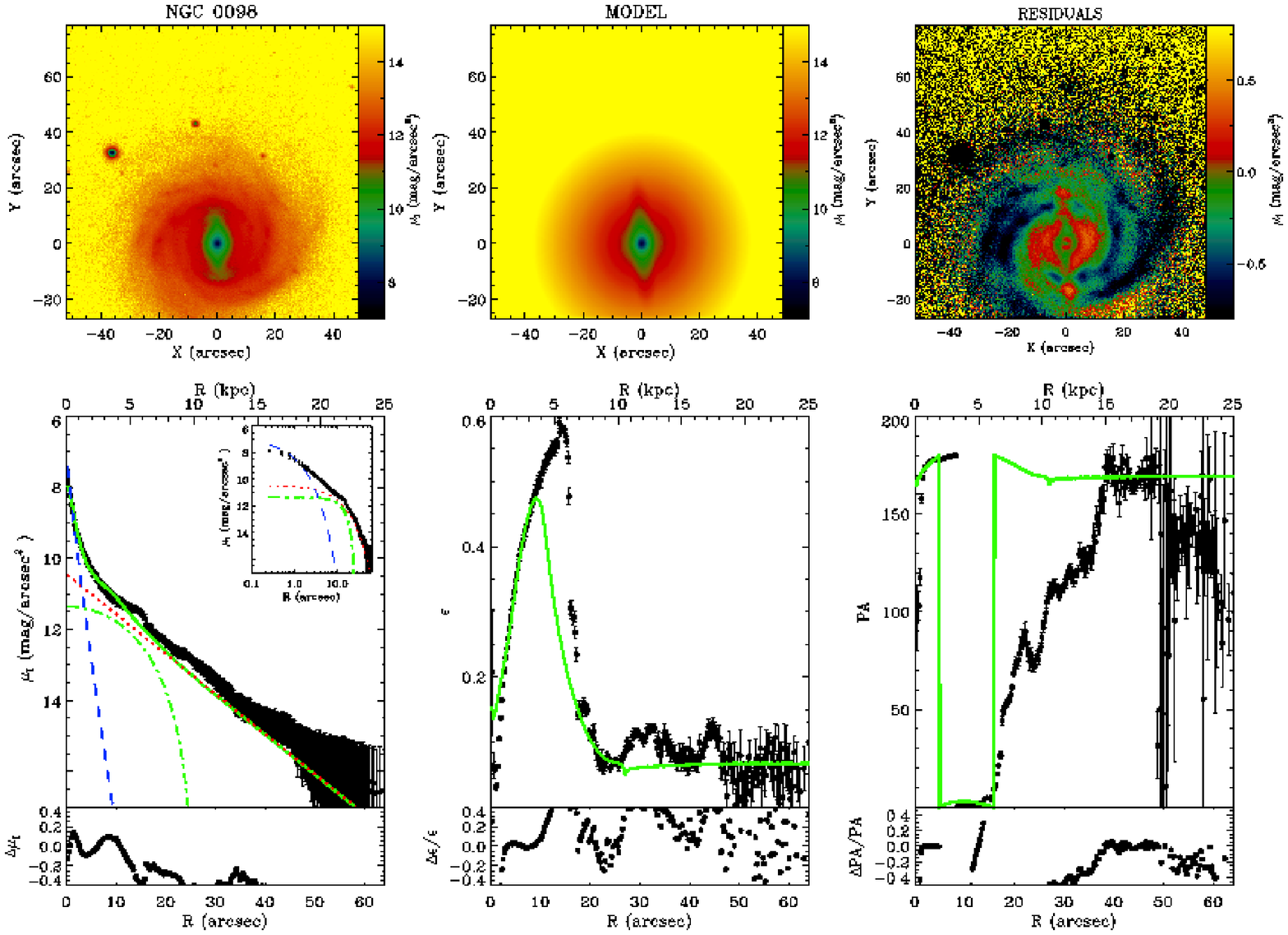}
   \includegraphics[angle=90,width=0.49\textwidth]{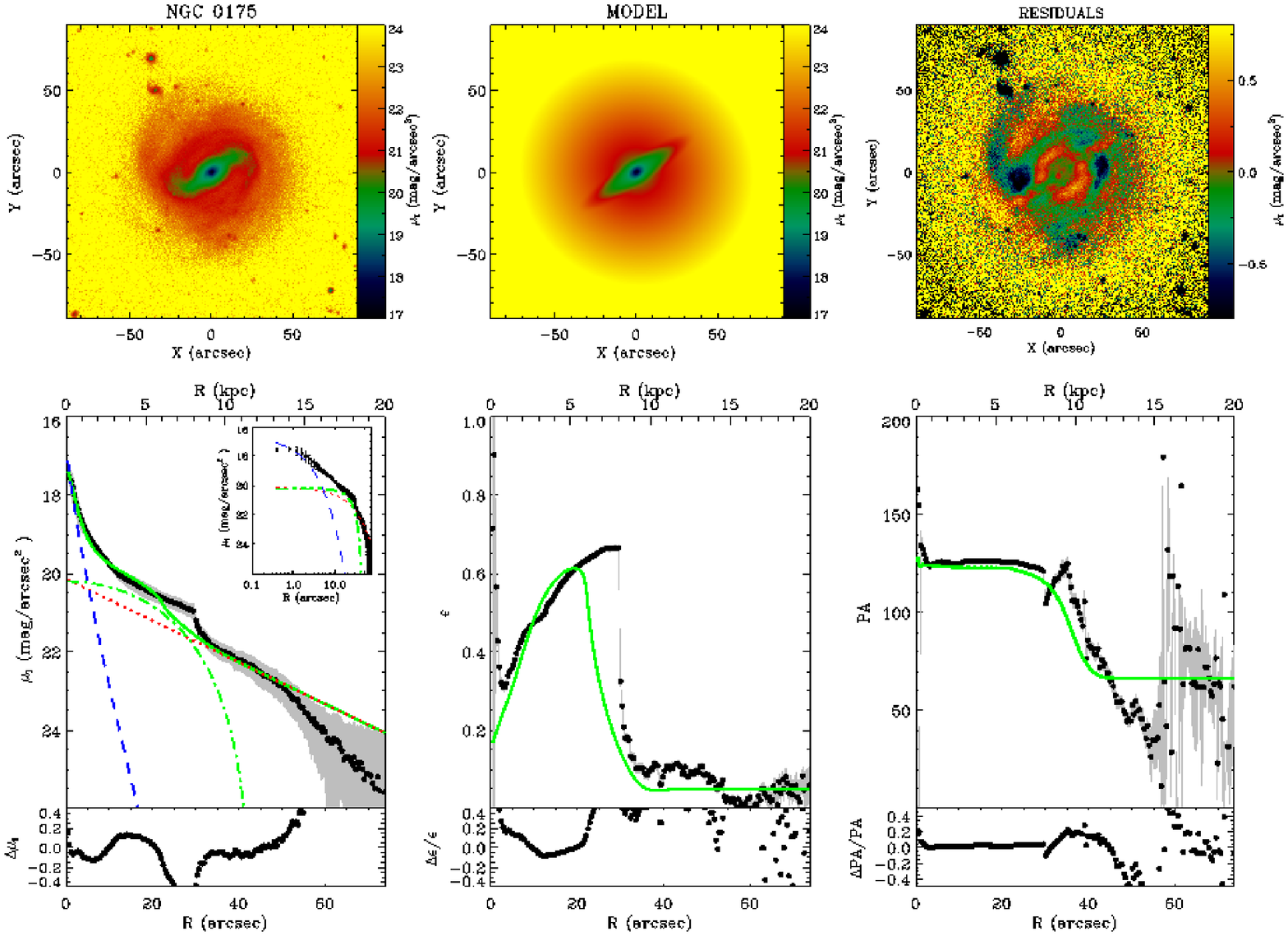}
   \includegraphics[angle=90,width=0.49\textwidth]{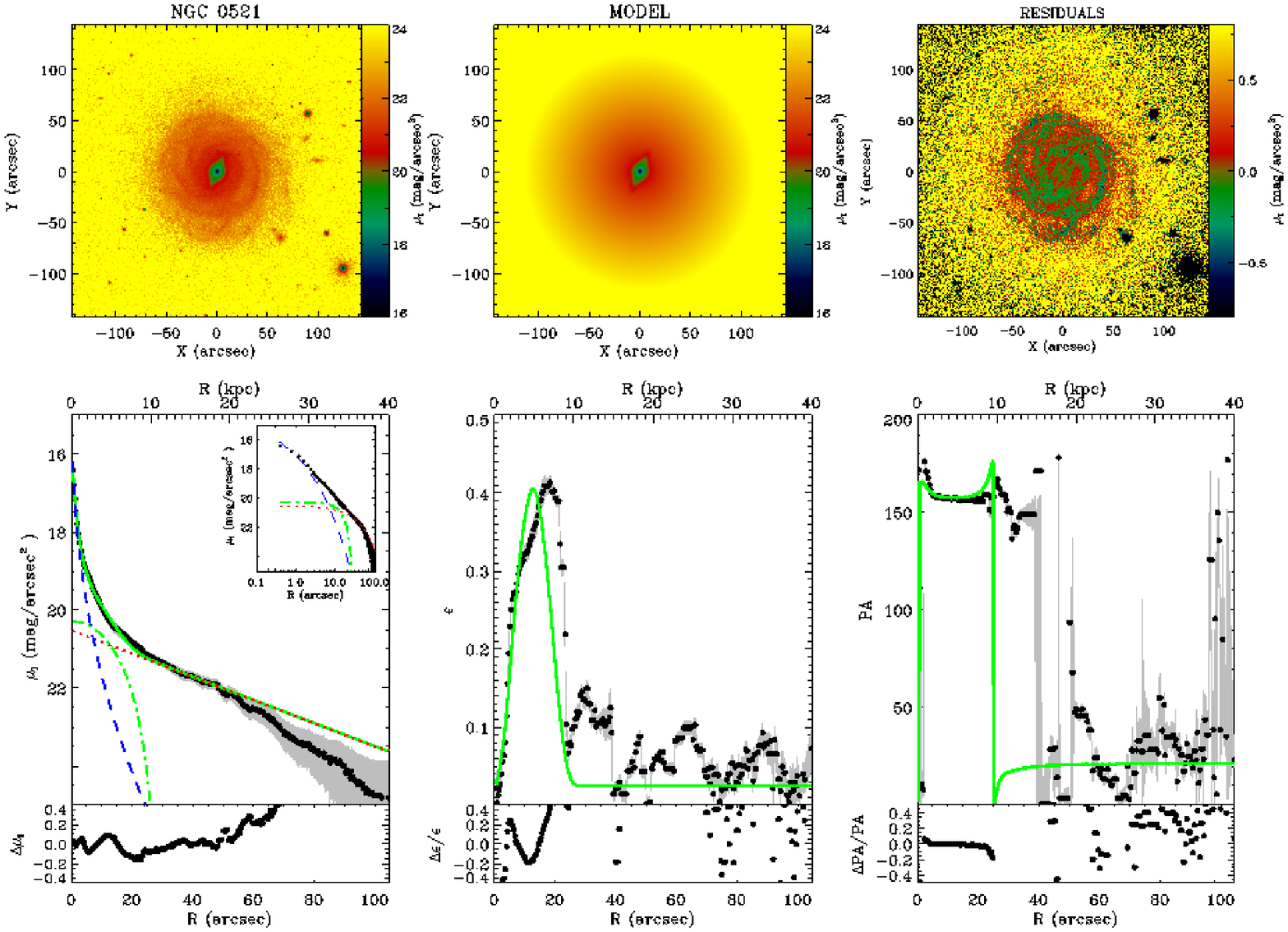}
   \includegraphics[angle=90,width=0.49\textwidth]{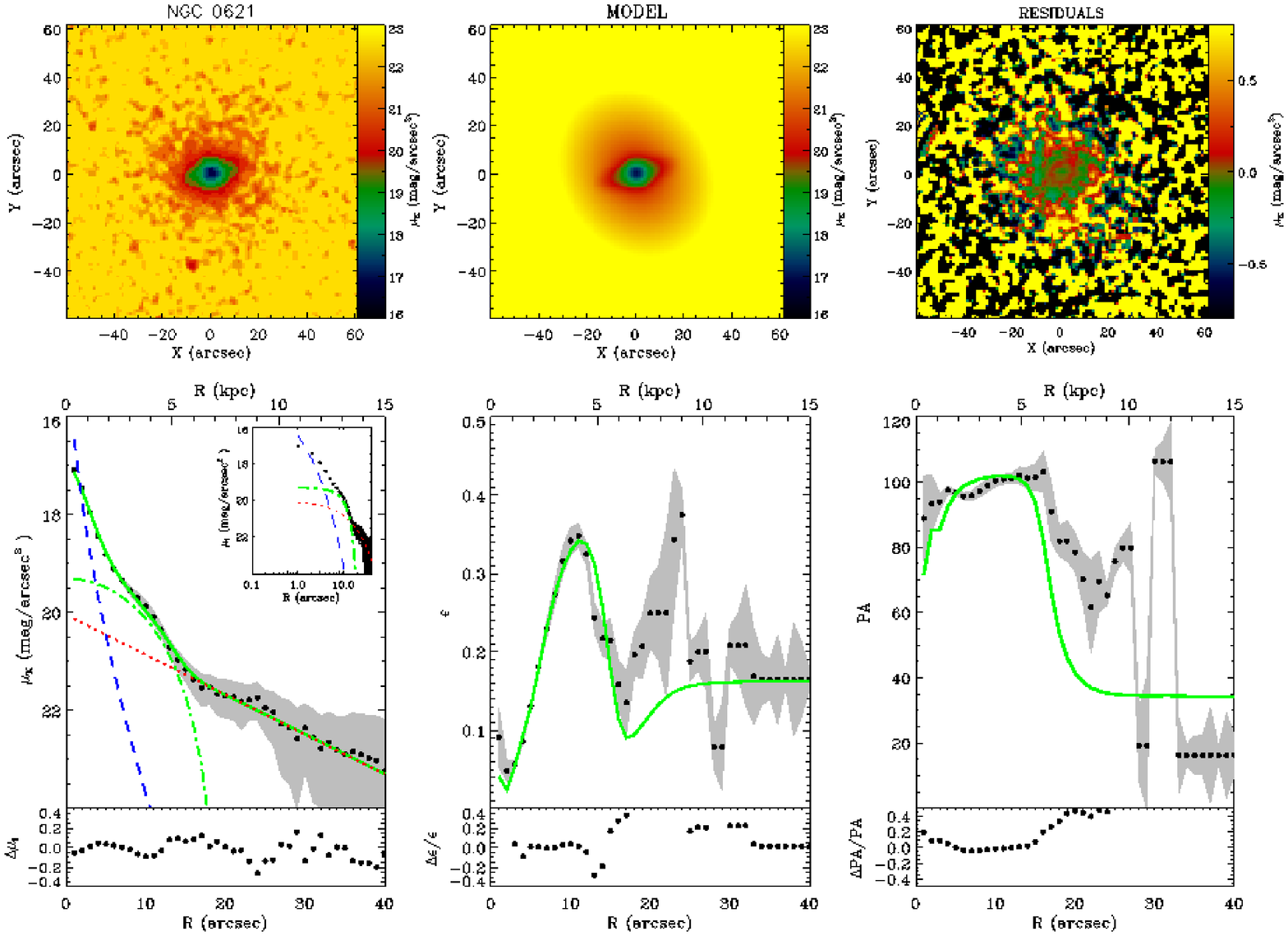}
      \caption{{\it  Top  left:}  Original  galaxy  image.   {\it  Top
          middle:} galaxy model derived  from GASP2D fit considering a
        bulge, a bar, and a disk component. {\it Top right:} residuals
        image derived  from the subtraction  of the galaxy  model from
        the original  image.  {\it Bottom left:}  ellipse averaged SBD
        of the  galaxy (black  dots).  Lines indicate  the fit  of the
        contribution  of  different  components derived  with  GASP2D:
        dashed line for the bulge, dotted-dashed line for the bar, and
        the dotted line  for the disk. The solid  green line indicates
        the ellipse  average SBD of  the best-fit model.   Upper inset
        shows  a zoom  of the  fit with  a logarithmic  scale for  the
        distance to  the center  of the  galaxy. {\it  Bottom middle:}
        ellipticity radial  profile measured  from the ellipse  fit on
        the galaxy  image (black  dots) and the  best fit  model image
        (green solid  line) {\it Bottom right:}  position angle ($PA$)
        radial profile  measured from  the ellipse  fit on  the galaxy
        image (black dots)  and the best fit model  image (green solid
        line). }
         \label{fig:phot}
   \end{figure*}
   \begin{figure*}
  \ContinuedFloat 
   \centering
   \includegraphics[angle=90,width=0.49\textwidth]{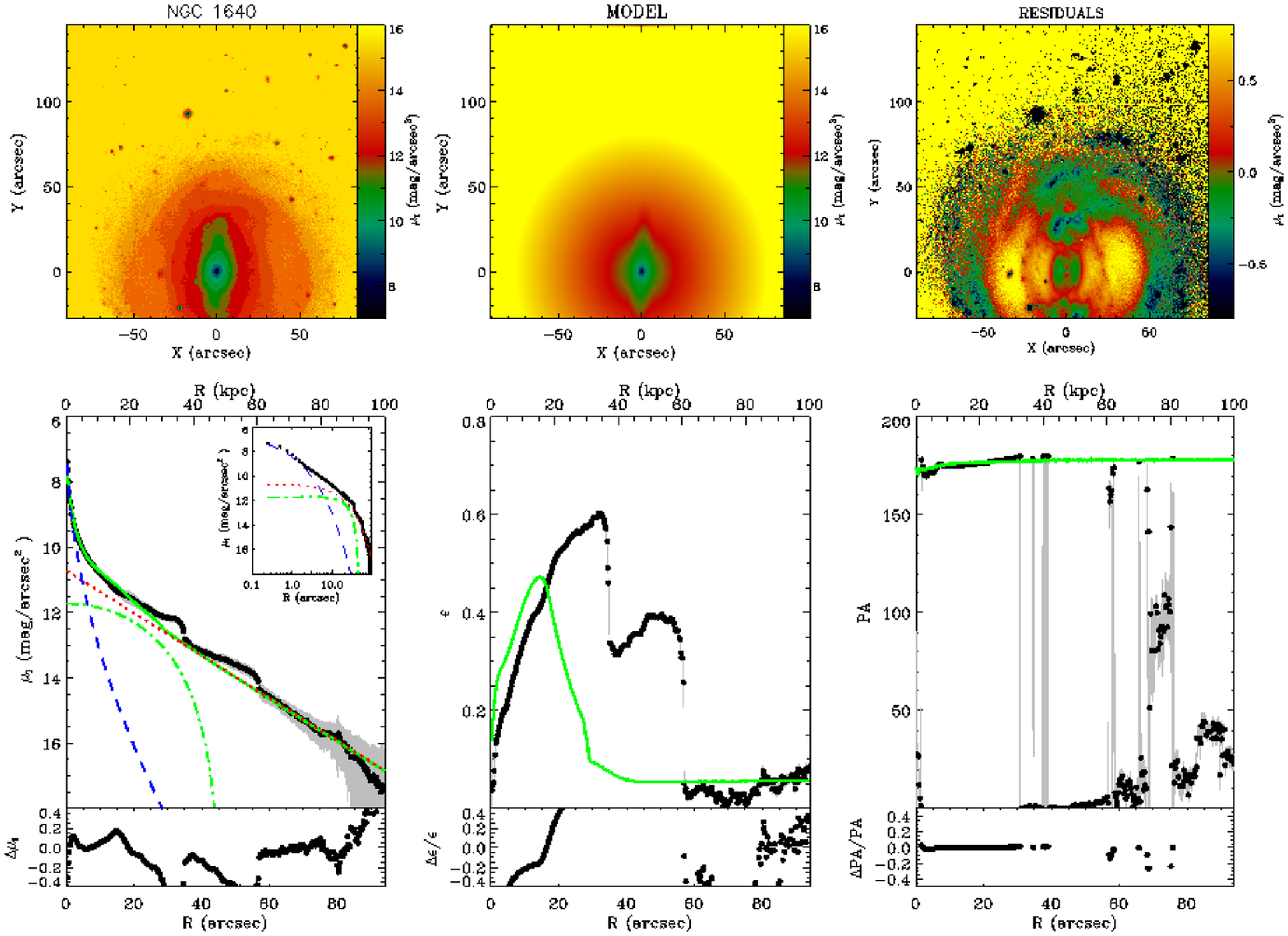}
   \includegraphics[angle=90,width=0.49\textwidth]{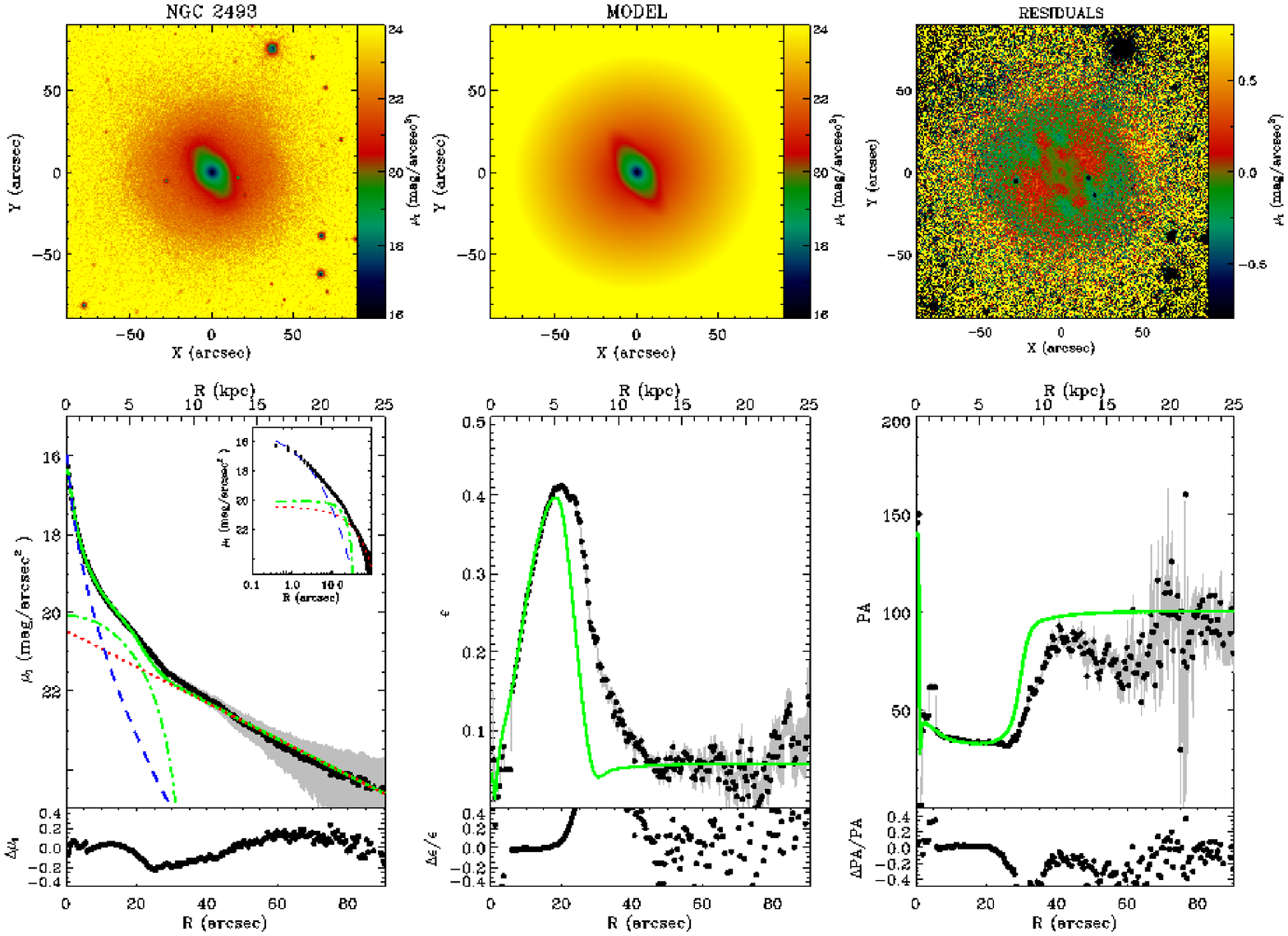}
   \includegraphics[angle=90,width=0.49\textwidth]{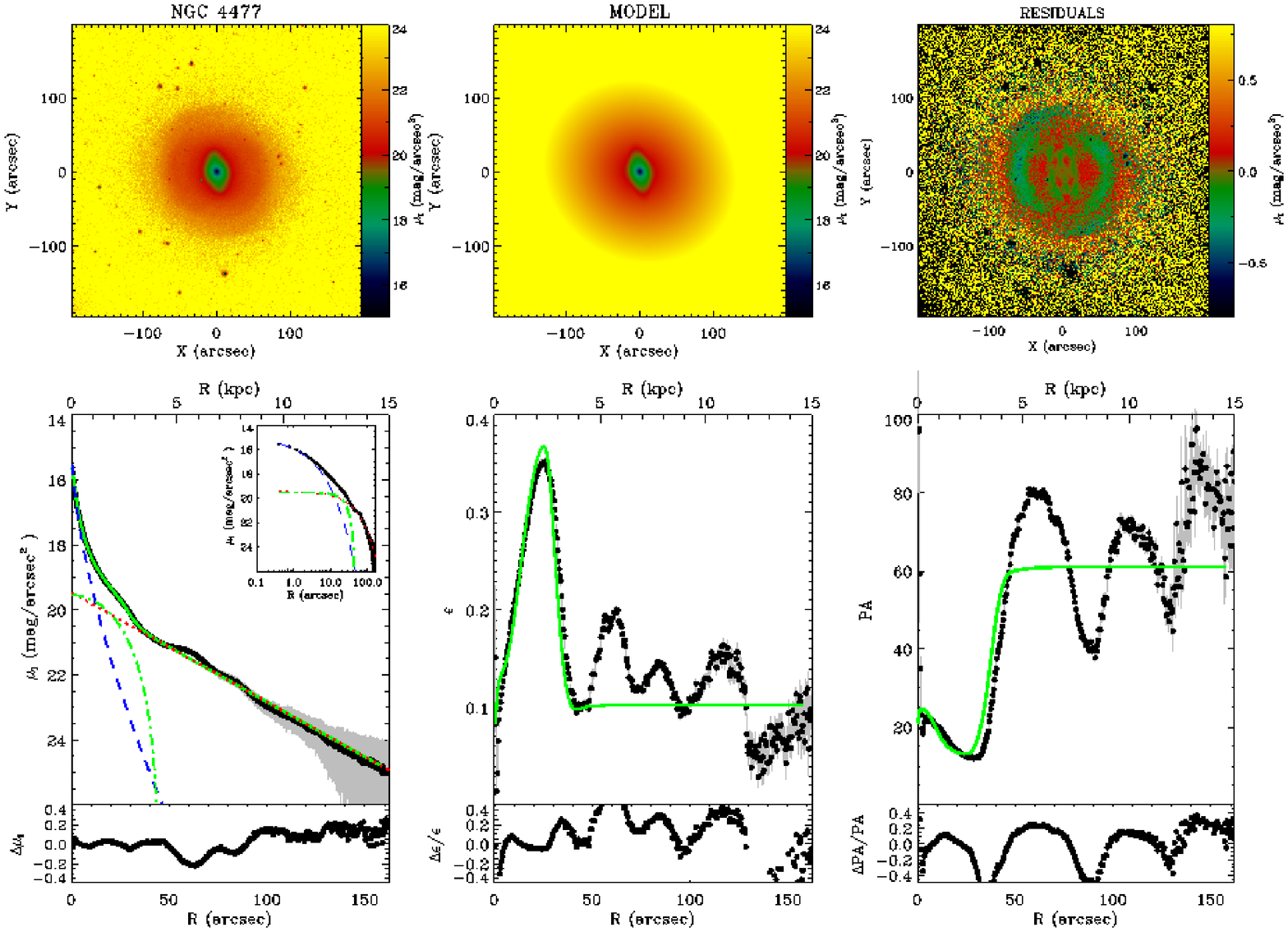}
   \includegraphics[angle=90,width=0.49\textwidth]{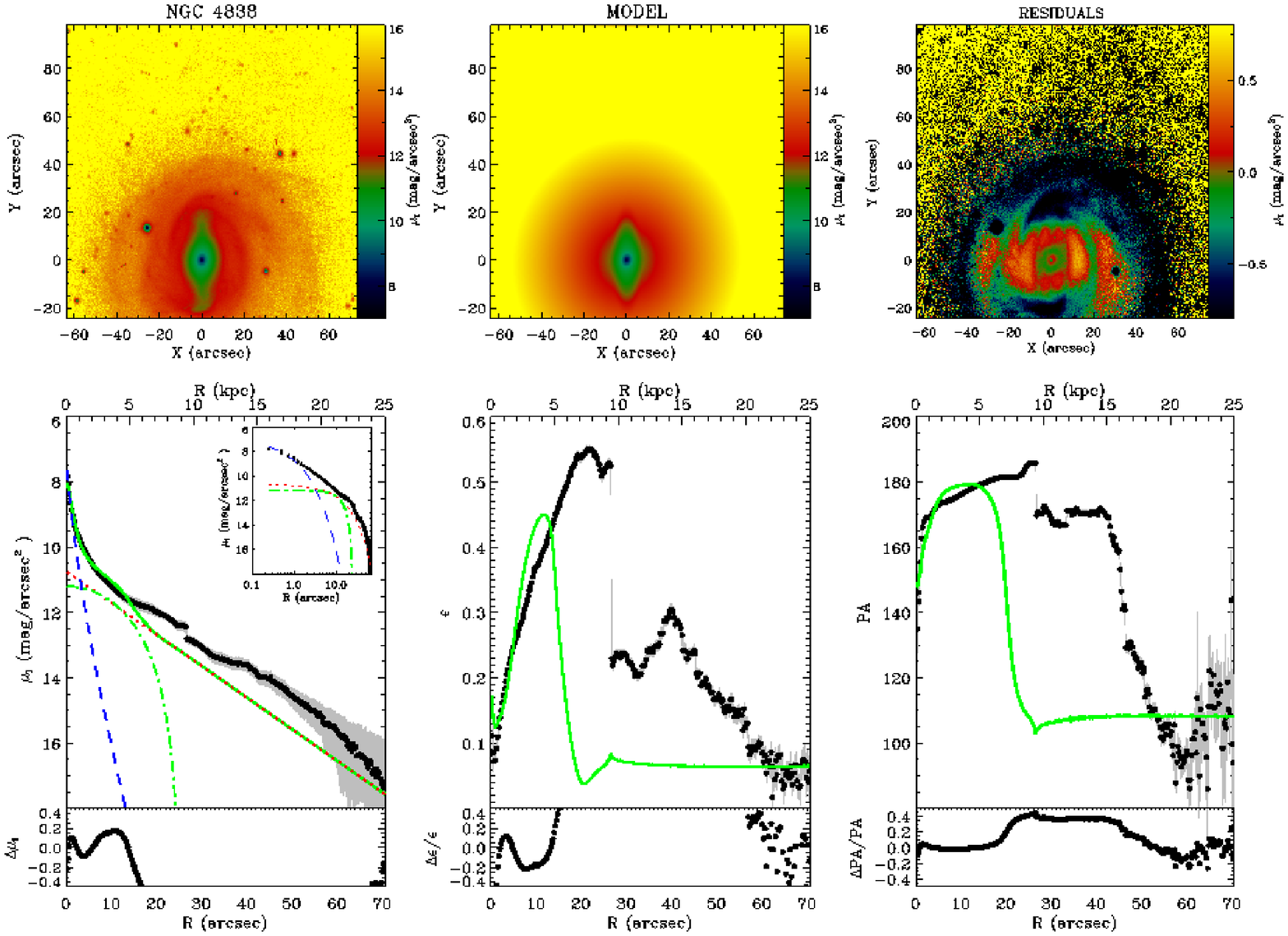}
      \caption{Continued.}
         \label{fig:phot}
   \end{figure*}

\section{Long-slit spectroscopy}
\label{sec:spectroscopy}

\subsection{Observations and data reduction}

The long-slit spectroscopic observations were performed using both the
VLT  at  the  European  Southern Observatory  (ESO)  and  Hobby-Eberly
Telescope (HET) at McDonald Observatory (Table \ref{tab:spec}). In all
cases the slit was placed along the major axis of the bar.

The VLT observations were carried  out in service mode using the Focal
Reducer   Low   Dispersion  Spectrograph   2   (FORS2)  mounting   the
volume-phased  holographic  grism   GRIS\_1028z$+$29  with  1028  $\rm
grooves\;mm^{-1}$  and  the  $0\farcs7\,\times\,6\farcm8$  slit.   The
detector was  a mosaic  of 2 MIT/LL  CCDs, each  with $2048\times4096$
pixels of $15\,\times\,15$ $\mu$m$^2$.  The wavelength range from 7681
\AA\ to 9423 \AA\ was covered in the reduced spectra with a reciprocal
dispersion  of  0.858  \AA\   pixel$^{-1}$  and  a  spatial  scale  of
$0\farcs250$ pixel$^{-1}$ after a $2\times2$ pixel binning.
Using  standard IRAF\footnote{IRAF  is distributed  by NOAO,  which is
  operated  by AURA  Inc., under  contract with  the National  Science
  Foundation.}   routines,  all  the  spectra  were  bias  subtracted,
flat-field  corrected,  cleaned  of  cosmic rays,  corrected  for  bad
pixels,  and wavelength  calibrated.  The  accuracy of  the wavelength
rebinning (1 km s$^{-1}$) was  checked by measuring the wavelengths of
the brightest  night-sky emission lines.  The  instrumental resolution
was $1.84\pm0.01$  \AA\ (FWHM)  corresponding to $\sigma_{\rm  inst} =
27$  km s$^{-1}$  at 8552  \AA .   The spectra  obtained for  the same
galaxy  were co-added  using the  center of  the stellar  continuum as
reference.   In  the  resulting   spectra  the  sky  contribution  was
determined by  interpolating along the outermost  $\approx30''$ at the
edges of the slit and then subtracted.

The  HET observations  were  carried  out in  service  mode using  the
Marcario Low  Resolution Spectrograph (LRS)  with the E2  phase volume
holographic  GRISM  and $1\farcs0$  slit.   The  detector was  a  Ford
Aerospace   3072$\times$1024  pixels   of  15$\times$15   $\mu$m$^{2}$
yielding a spatial scale  of 0$\farcs$235 pixel$^{-1}$.  The wavelength
range covered  from 4790 \AA\ to  5850 \AA\ and was  sampled with 0.38
\AA\ pixel$^{-1}$.
We used  standard IRAF  routines to reduce  the long-slit  spectra. In
addition to  the bias subtraction,  flat field correction,  and cosmic
ray rejection, we corrected for a two degree tilt between the slit and
the CCD  rows by performing  a subpixel  shifting of the  CCD columns.
After this  correction, we performed the  wavelength calibration using
neon  arc frames.   The accuracy  of  the wavelength  rebinning (1  km
s$^{-1}$) was  checked by measuring  the wavelengths of  the brightest
night-sky   emission   lines.    The   instrumental   resolution   was
$1.73\pm0.01$ \AA\ (FWHM)  corresponding to $\sigma_{\rm inst}  = 42 $
km s$^{-1}$ at 5200 \AA.   No correction for anamorphic distortion was
performed on our spectra due to a lack of suitable stars to trace this
effect. Based  on Fig. 2 in  \citet{saglia10}, variations of up  to 10
pixels are expected in the red and blue extremes of the CCD \citep[see
  also][]{fabricius12}.   However, for  the  central wavelengths,  the
anamorphic distortion is negligible for  every spatial position in the
CCD.  Since we are interested  in measuring the Mg~{\small I} triplet,
we perform the fit using only a small spectral region thus avoiding as
much as  possible this effect.   In addition, possible  variations are
expected to be well below the typical spatial bin sizes that we use at
the ends of the slit. To  correct for flexure of the instrument during
the night we measured the  wavelength position of the 5577\AA\ skyline
and correct  the wavelength  calibration to zeroth  order by  adding a
constant offset.  The median absolute  offset of all observations is 5
km s$^{-1}$ .

\begin{table}
\caption{Characteristics of the spectroscopic observations}   
\label{tab:spec}    
\centering            
\begin{tabular}{c c c c c c}      
\hline\hline              
Galaxy      &  Telesc.     &Instr.    & $\sigma_{\rm intr}$ & PA         & Exp Time \\
            &              &          & (km s$^{-1}$)      & ($^{\circ}$)   &  (min)   \\  
  (1)       &      (2)     &   (3)    &  (4)              &  (5)       & (6)   \\
\hline
   IC~1815  &  HET         & LRS      & 42                & 131  & 1$\times$30 \\
   NGC~0043 &  HET         & LRS      & 42                & 98   & 3$\times$30 \\
   NGC~0098 &  VLT         & FORS2    & 27                & 32   & 4$\times$45 \\
   NGC~0175 &  VLT         & FORS2    & 27                & 125  & 4$\times$45 \\
   NGC~0521 &  VLT         & FORS2    & 27                & 156  & 4$\times$45 \\
   NGC~0621 &  HET         & LRS      & 42                & 100  & 2$\times$30 \\
   NGC~1640 &  VLT         & FORS2    & 27                & 47   & 4$\times$45 \\
   NGC~2493 &  HET         & LRS      & 42                & 32   & 4$\times$45 \\
   NGC~4477 &  VLT         & FORS2    & 27                & 13   & 1$\times$45 \\
   NGC~4838 &  VLT         & FORS2    & 27                & 52   & 4$\times$45 \\
\hline                                 
\end{tabular}
\tablefoot{(1) Galaxy name; (2) and  (3) telescope and instrument used
  in the observations; (4)  instrumental resolution; (5) position angle
  of the slit; (6) exposure time of the observations.}
\end{table}

\subsection{Measuring stellar kinematics}
\label{sec:stellar}

The stellar kinematics  of the sample galaxies were  measured from the
galaxy absorption features present in the wavelength range centered on
either the  Ca~{\small II}  triplet ($\lambda\lambda$8498,  8542, 8662
\AA)   for   the  VLT   spectra   or   Mg~{\small  I}   line   triplet
($\lambda\lambda$5164, 5173, 5184  \AA) for the HET  spectra using the
penalized pixel-fitting  method \citep[pPXF;][]{cappellariemsellem04}.
The  spectra  were  rebinned  along  the  dispersion  direction  to  a
logarithmic  scale,  and  along  the spatial  direction  to  obtain  a
signal-to-noise ratio S/N=20 per resolution bin at the outer radii. It
increases to S/N $>$ 50 at the center of the galaxies.

A  linear   combination  of  stellar  templates   convolved  with  the
line-of-sight   velocity  distribution   (LOSVD)   described  by   the
Gauss-Hermite expansion \citep{vandermarelfranx93}  was fitted to each
galaxy spectrum by  $\chi^{2}$ minimization in the  pixel space.  This
allowed us  to derive  radial profiles of  the mean  velocity ($v_{\rm
  los}$), velocity dispersion ($\sigma_{\rm los}$), and third- ($h_3$)
and fourth-order ($h_4$) Gauss-Hermite moments.  The stellar templates
used  in this  study were  obtained  from the  Ca~{\small II}  triplet
library  \citep{cenarro01}, which  cover  the  spectral range  between
8350-9020  \AA\   at  1.5  \AA\   (FWHM),  and  the   INDO-US  library
\citep{valdes04}, which covers the  wavelength range 3460-9464 \AA\ at
$\sim$1.36   \AA\    (FWHM)   for    the   VLT   and    HET   spectra,
respectively. Prior to fitting, they were convolved with a Gaussian to
match the  instrumental resolution of our  spectra.  The uncertainties
on the kinematic parameters were  estimated by Monte Carlo simulations
including photon, read-out, and sky noise.

Figure  \ref{fig:kin}  shows the  kinematic  radial  profiles for  the
sample galaxies. It is worth remembering here that the radial profiles
obtained in  this paper  are along  the major axis  of the  bars. This
implies  that  radial velocities  should  be  corrected not  only  for
inclination, but also for the different position angle with respect to
the  line-of-nodes (LON).   In addition,  galaxies in  the sample  are
nearly face-on, thus our LOS samples mostly the vertical components of
the velocity and velocity dispersion.

   \begin{figure*}[!ht]
   \centering
   \includegraphics[bb=150 400 450 900,width=6cm]{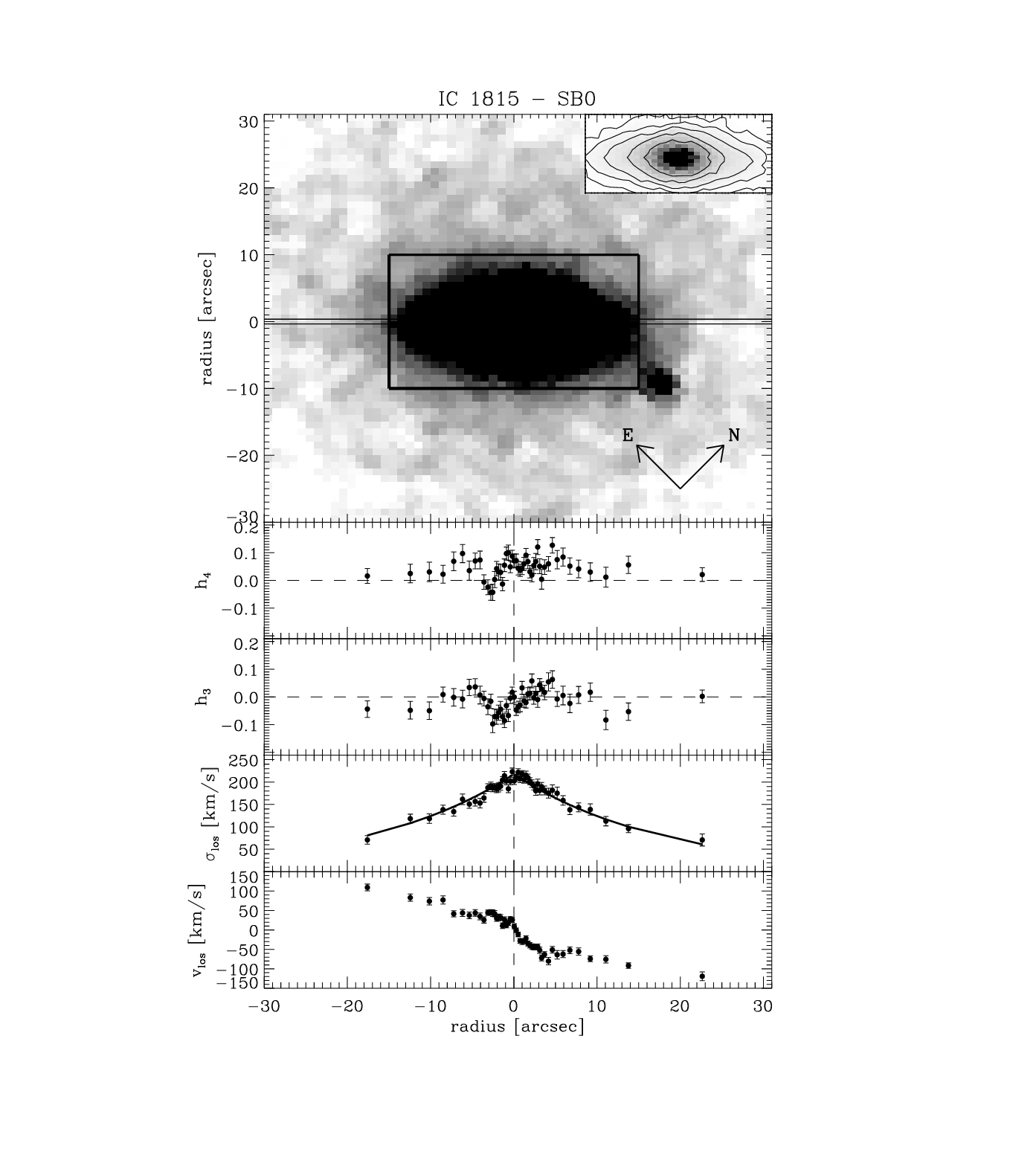}
   \includegraphics[bb=150 400 450 900,width=6cm]{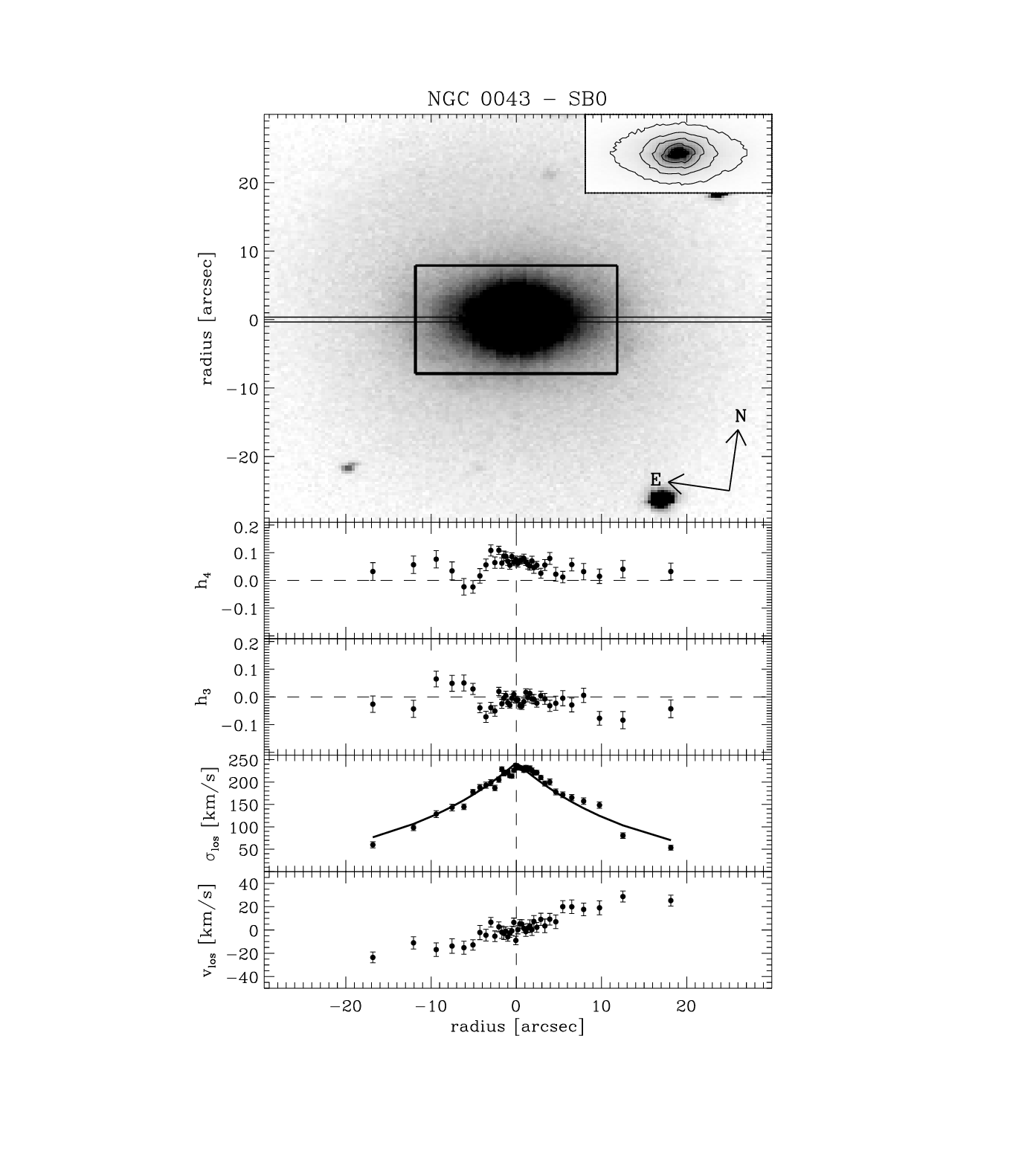}
   \includegraphics[bb=150 400 450 900,width=6cm]{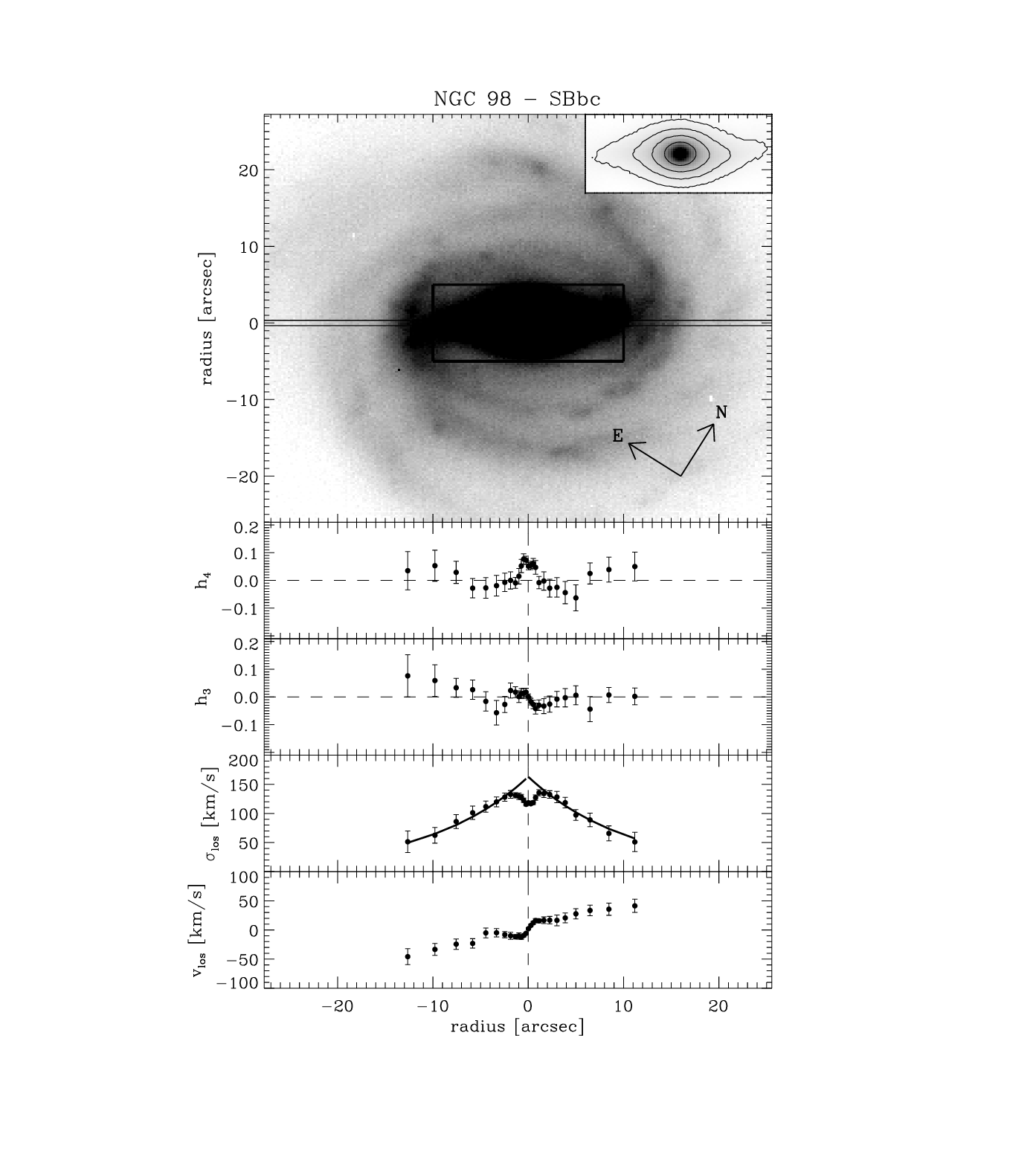}
   \includegraphics[bb=150 400 450 900,width=6cm]{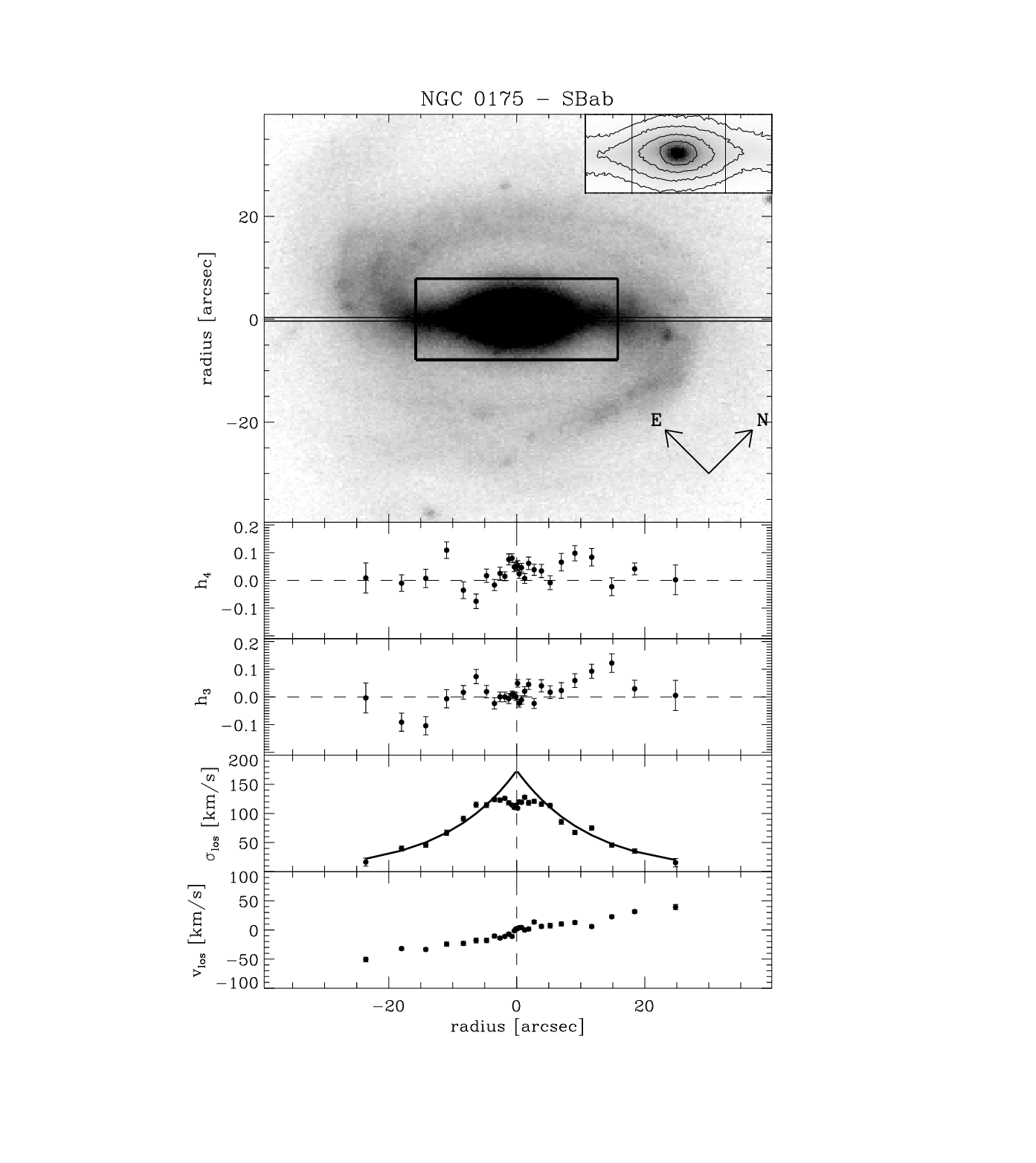}
      \caption{Morphology and stellar kinematics of the sample galaxies.  
 For  each galaxy  the top  panel shows  the galaxy  image.  The  slit
 position and  image orientation  are indicated.  The inset  shows the
 portion of the  galaxy image marked with a black  box. The gray scale
 and isophotes  were chosen  to enhance the  features observed  in the
 central regions.
 The remaining panels  show from top to bottom the  radial profiles of
 $h_4$,  $h_3$,  $\sigma_{\rm los}$,  and  velocity  $v$ (obtained  by
 subtracting  the systemic  velocity  from $v_{\rm  los}$).  The  best
 exponential  fit to  the $\sigma_{\rm  los}$ radial  profile is  also
 shown by a black solid line in the corresponding panel.}
         \label{fig:kin}
   \end{figure*}
   \begin{figure*}
  \ContinuedFloat
   \centering
   \includegraphics[bb=150 400 450 900,width=6cm]{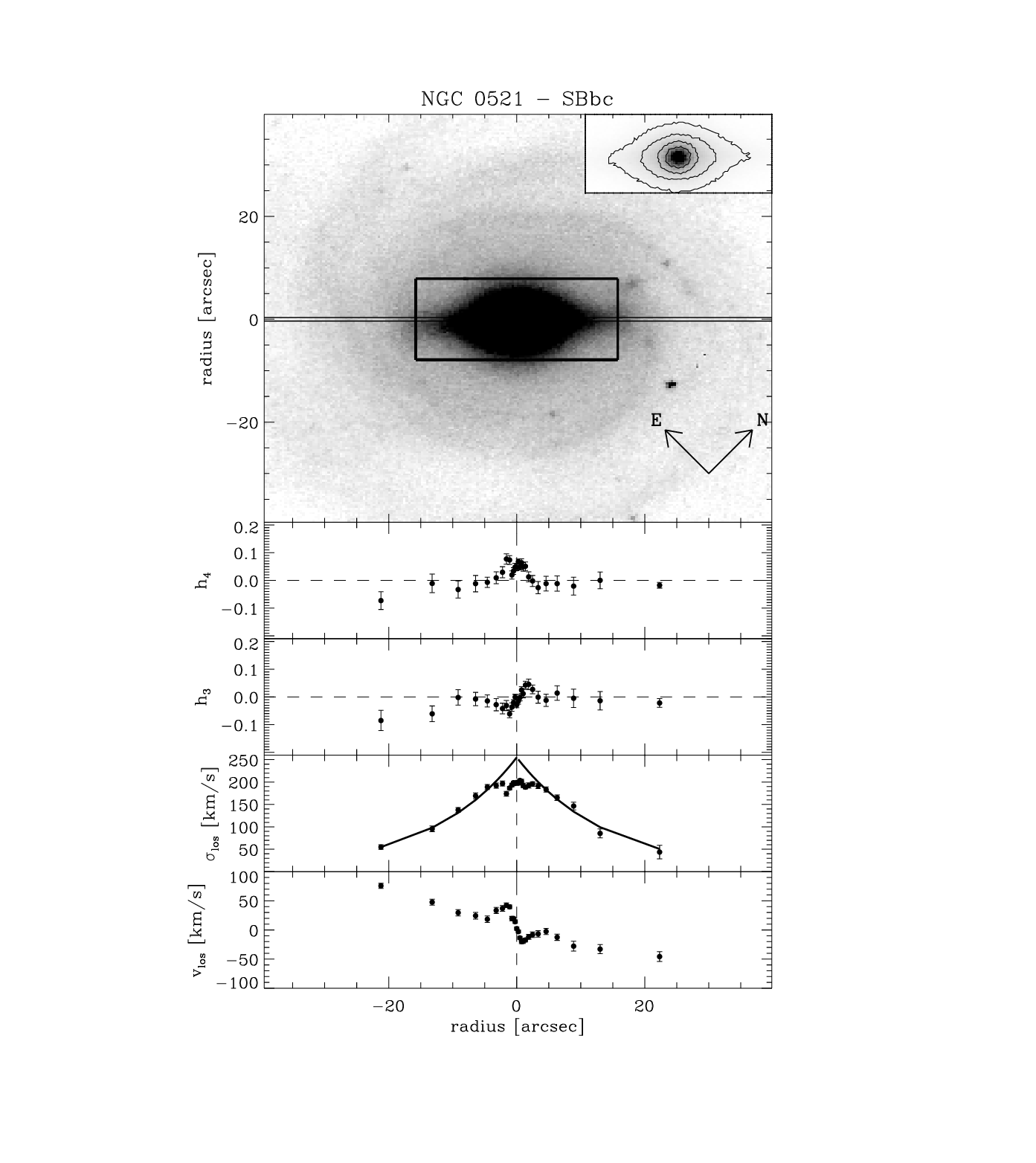}
   \includegraphics[bb=150 400 450 900,width=6cm]{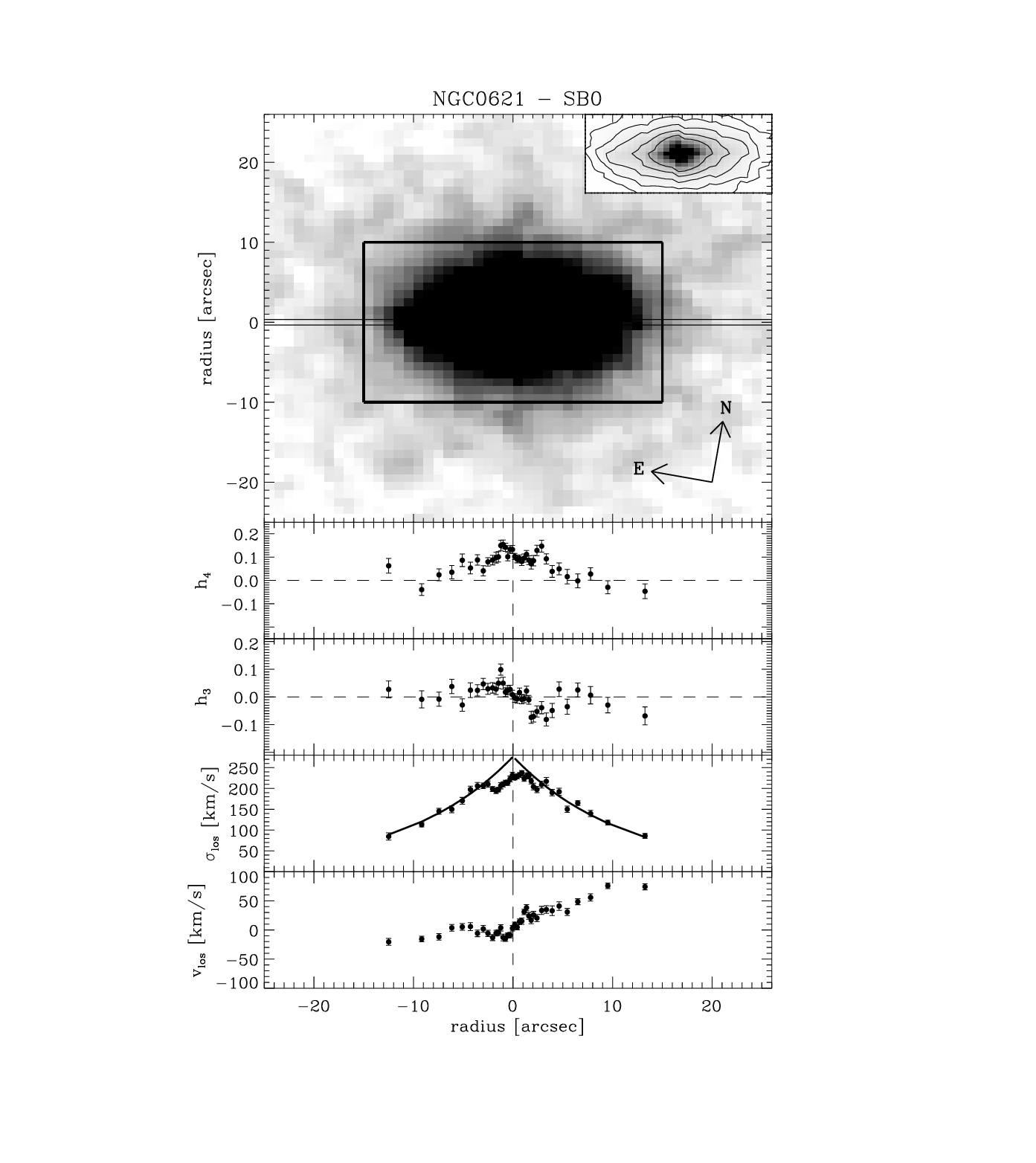}
   \includegraphics[bb=150 400 450 900,width=6cm]{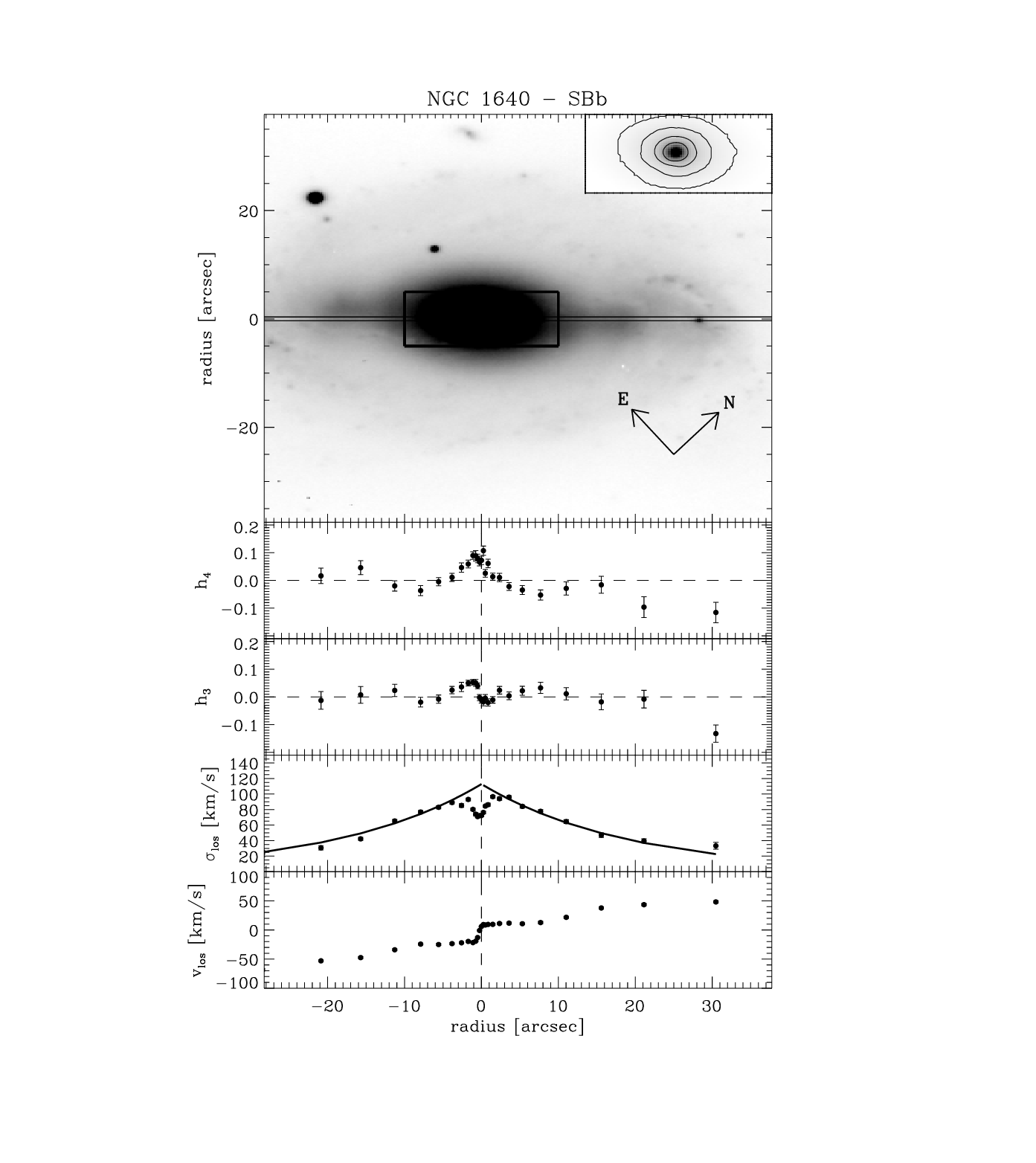}
   \includegraphics[bb=150 400 450 900,width=6cm]{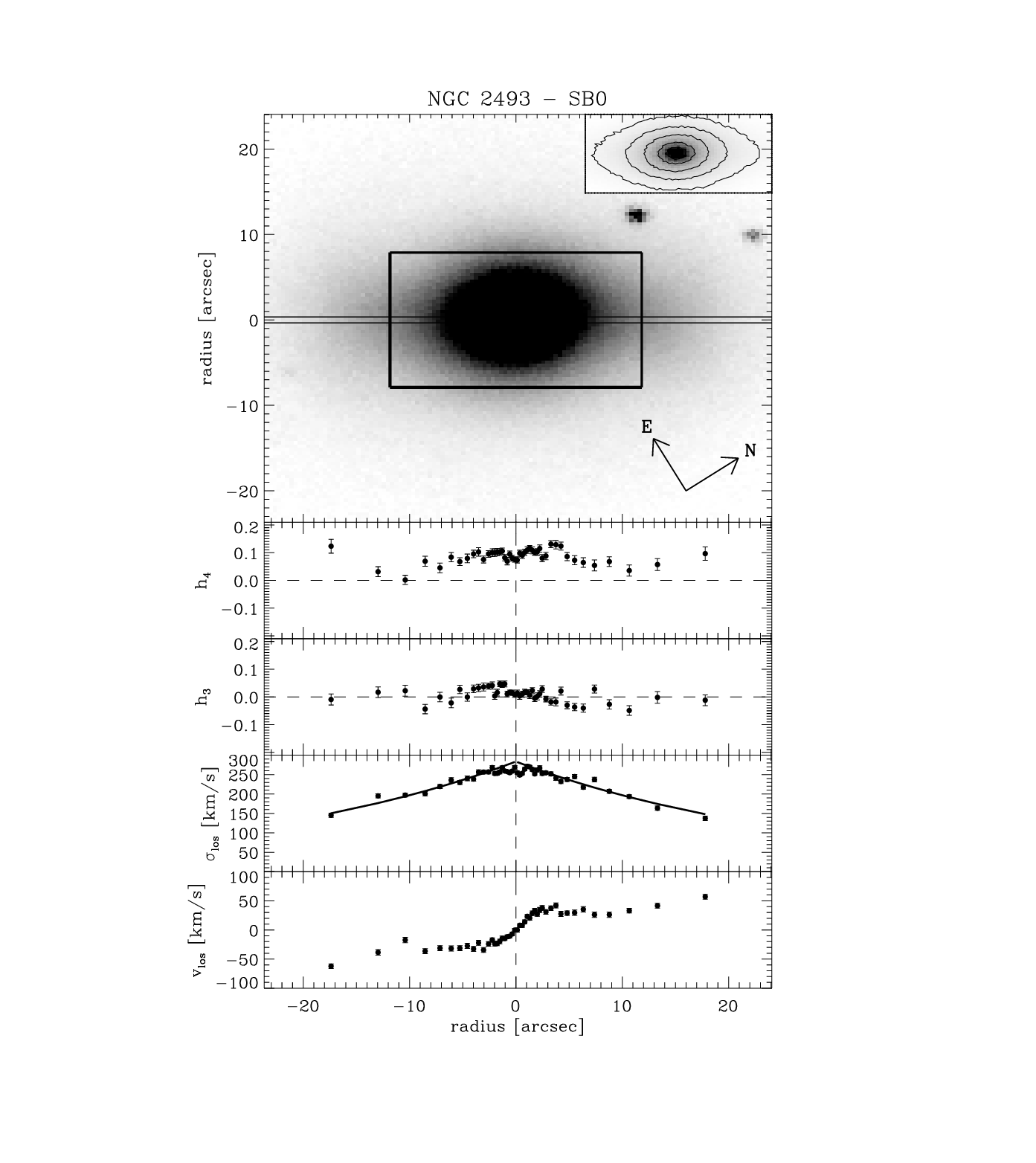}
   \end{figure*}
   \begin{figure*}
  \ContinuedFloat 
   \centering
   \includegraphics[bb=150 400 450 900,width=6cm]{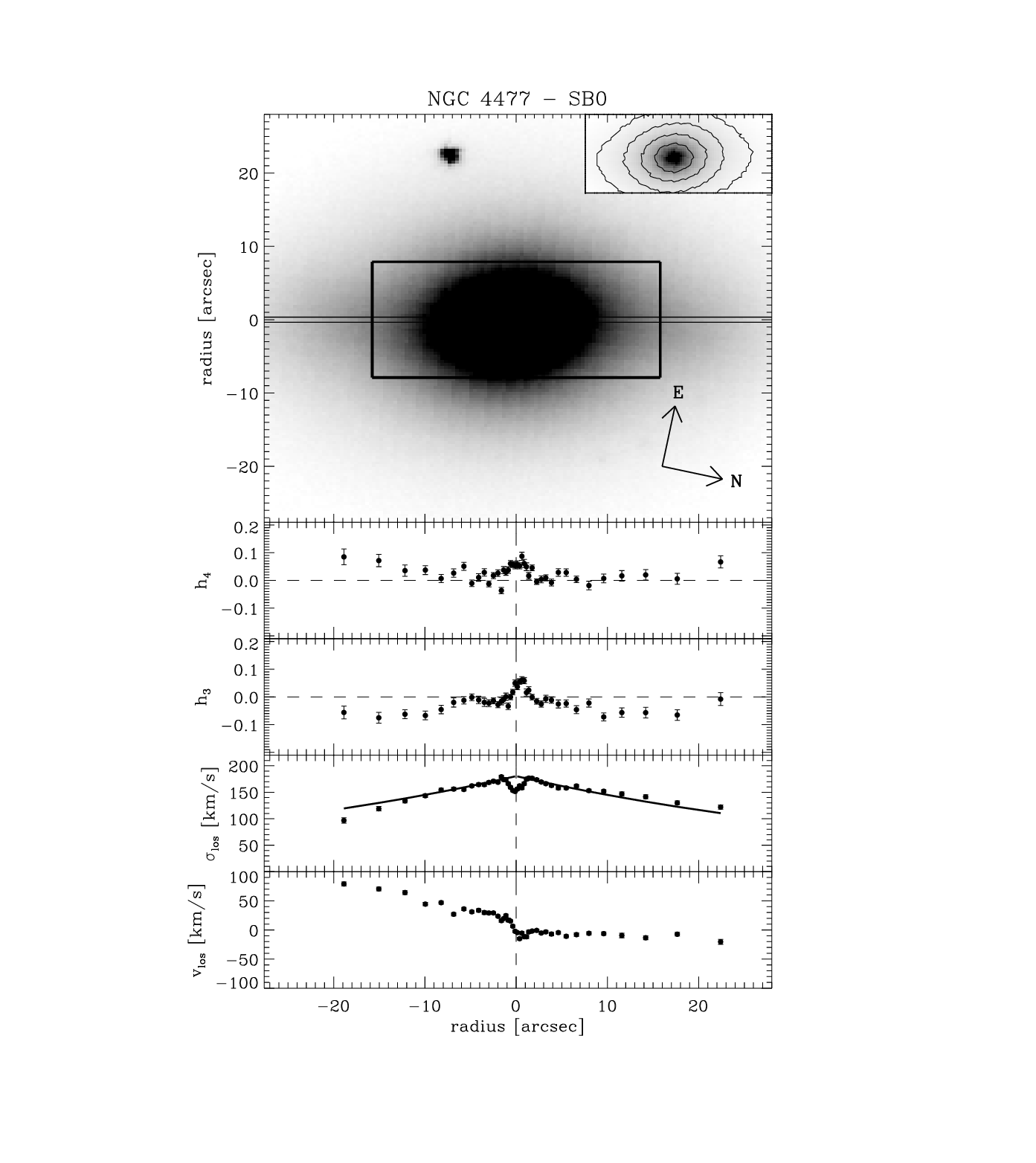}
   \includegraphics[bb=150 400 450 900,width=6cm]{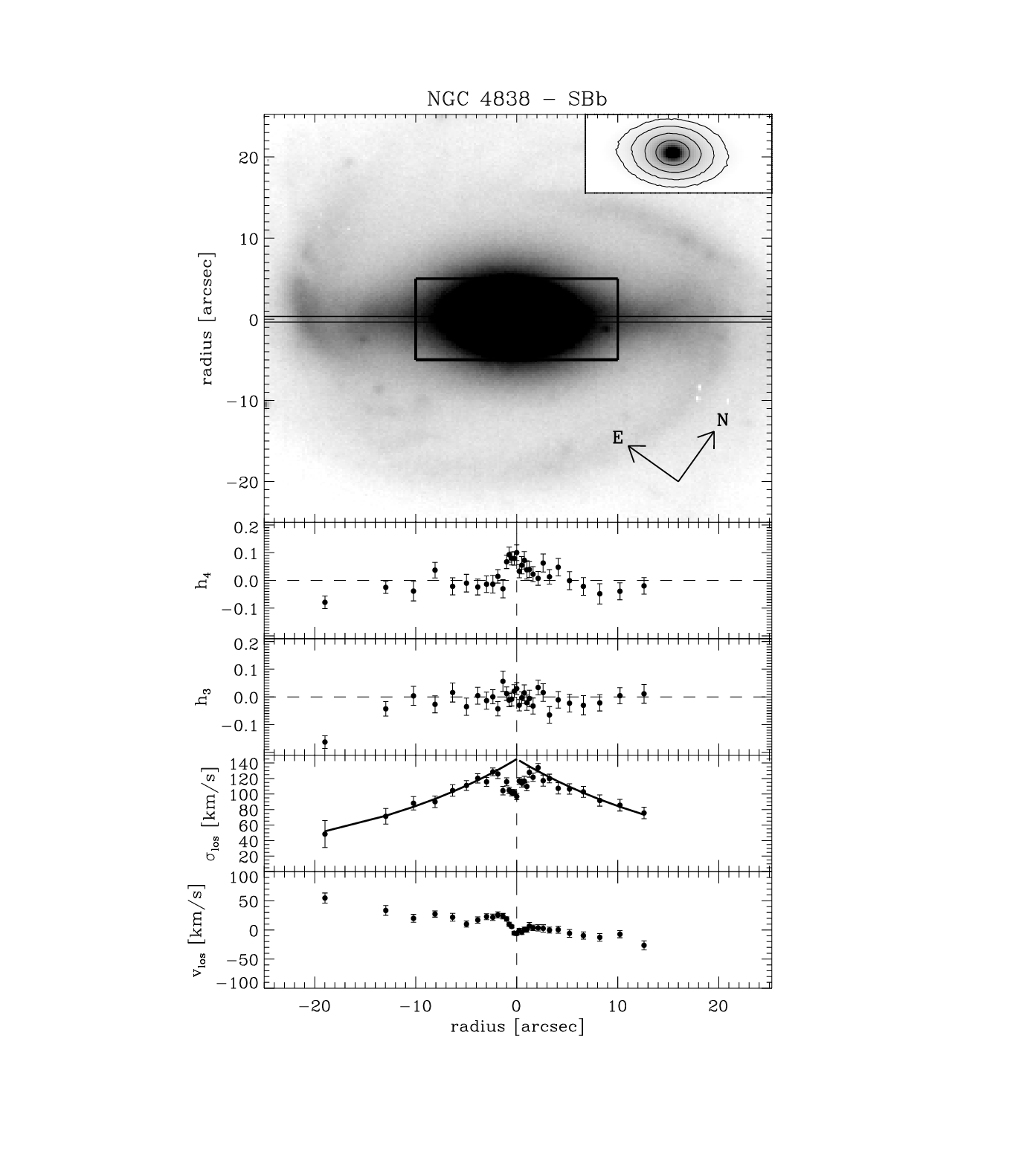}
      \caption{Continued.}
   \end{figure*}

\section{Results}
\label{sec:results}

\subsection{Photometric properties of the sample}
\begin{figure}[!t]
   \centering
   \includegraphics[width=0.49\textwidth]{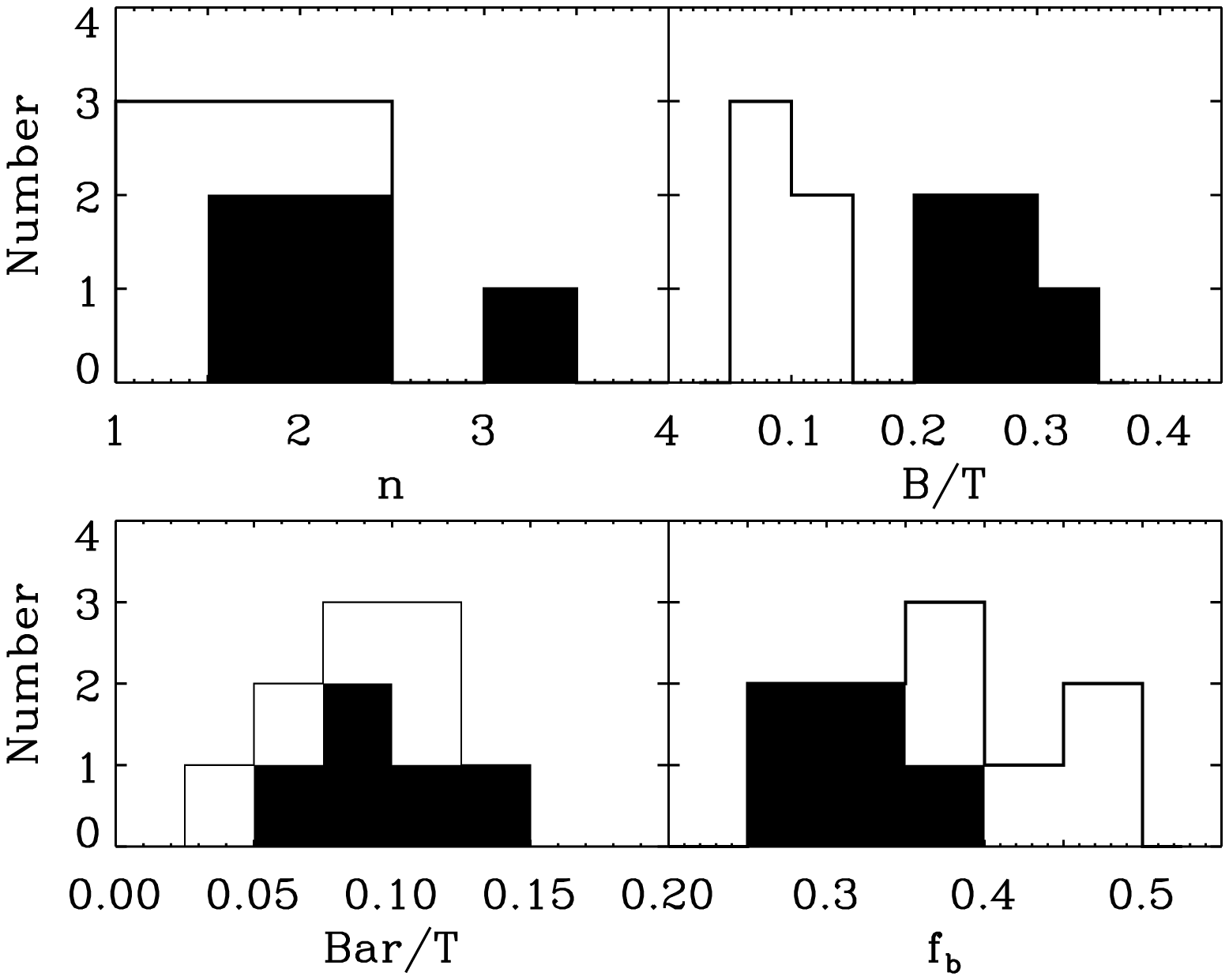}
      \caption{Distribution   of    the   $n$    S\'ersic   parameter,
        bulge-to-total   ($B/T$)    luminosity   ratio,   bar-to-total
        ($Bar/T$) luminosity ratio, and bar strength ($f_{\rm b}$) for
        our   sample  galaxies.    Filled  histograms   represent  our
        subsample of lenticular galaxies.}
         \label{fig:photpar}
   \end{figure}

Figure  \ref{fig:photpar} shows  the distribution  of S\'ersic  index,
bulge-to-total ($B/T$)  and bar-to-total ($Bar/T$)  luminosity ratios,
and bar strength.

The S\'ersic  index distribution  for our  sample galaxies  covers the
range $1.2\leq n  \leq3.4$ with six of them having  $n<2$.  We did not
find any correlation between S\'ersic  index and Hubble type. We found
a  large spread  of S\'ersic  index ranging  from $1.6<n<3.4$  for our
sample of  5 lenticular  galaxies, whereas  the remaining  spirals all
have  $n<2.4$.  This behaviour  for  the  lenticular galaxies  is  not
surprising,  and   has  been  already  noticed   in  previous  studies
\citep{mendezabreu08a, laurikainen10}.   The low  $n$ values  found in
our galaxies can  be explained by the inclusion of  a bar component in
our       photometric       decompositions.        Several       works
\citep[][]{aguerri05,balcells07,gadotti08,laurikainen10}   have  shown
how the presence of another component  (mainly a bar) at the center of
the  galaxy,  and  properly  taken into  account  in  the  photometric
decomposition, reduces the $n$ index of the bulge.

The effect of including a bar in the photometric decomposition is also
clearly shown  in the  $B/T$ distribution. Our  sample spans  only the
range between $0.07<B/T<0.33$  even if it is  biased toward early-type
disk galaxies.  \citet{gadotti08} estimated that not including the bar
in the  photometric decomposition can  result in an  overestimation of
the $B/T$ ratio by  a factor of 2.  Our $B/T$  distribution is also in
agreement with other results \citep{aguerri05,laurikainen07}.

The   bar-to-total    ($Bar/T$)   ratio   covers   the    range   from
$0.02<Bar/T<0.14$ indicating  that bars  contribute marginally  to the
total  luminosity, and  possibly  to  the total  mass  of the  galaxy.
Nonetheless, they  are able to  change the morphology and  dynamics of
their central regions. The bar  strength distribution spans from $0.26
\leq f_{\rm b}  \leq 0.46$. We found that bars  in lenticular galaxies
are weaker than in spirals. Similar results have been already found in
the literature \citep[i.e.,][]{aguerri09} but  our sample is too small
to reach more statistically meaningful conclusions.

\subsection{$\sigma-$drops}
\label{sec:stellar}

Central reductions of the stellar  velocity dispersion of galaxies are
usually called $\sigma-$drops \citep{emsellem01}. Figure \ref{fig:kin}
shows clearly the ubiquity  of $\sigma-$drops and $\sigma-$plateaus in
our  galaxy sample.  We found  a clear  drop in  the central  velocity
dispersion of five galaxies and another  three show a plateau in their
central regions.  This  means that $80\%$ of our  sample galaxies show
this  kinematic feature  in their  centers.  This  is consistent  with
previous results for barred galaxies \citep{perez09}.

In order  to characterize  the $\sigma-$drop  properties we  fitted an
exponential  profile  to  the  velocity  dispersion  profile  of  each
galaxy. In  the fit,  we excluded  the central  region of  each galaxy
where the  drop/plateau appears.   Both sides  of the  radial profiles
were fitted simultaneously  and the errors on  the velocity dispersion
were used as weights into the fit.  The results of these fits are also
plotted  in  Fig.~\ref{fig:kin}  and   the  values  of  the  resulting
scale-lengths   are   quoted   as  $\sigma_{\rm   scale}$   in   Table
\ref{tab:kinproperties}.  For each galaxy  where a $\sigma-$drop, or a
$\sigma-$plateau,  is present  we measured:  the drop  radius, as  the
radius  where  the  velocity   dispersion  profile  departs  from  the
exponential fit;  the maximum and  minimum of the  velocity dispersion
within this radius; and  the integrated velocity dispersion difference
between the best fit and actual radial profile. These values, together
with     other     kinematic     properties,     are     listed     in
Table~\ref{tab:kinproperties}.

\begin{table*}
\caption{Kinematic properties of the sample galaxies}   
\label{tab:kinproperties}    
\centering            
\begin{tabular}{c c c c c c c c c}      
\hline\hline              
Galaxy      & $\sigma_0$     &  $\sigma_{\rm 0,model}$ & $\sigma_{r_{\rm e}/8}$ & $\sigma_{r_{\rm e}/8,{\rm model}}$ & $r_{\rm drop}$  & $\sigma_{\rm max,drop}$& $\sigma_{\rm int}$ & $\sigma_{\rm scale}$ \\
            &  (km s$^{-1}$) &  (km s$^{-1}$)         & (km s$^{-1}$)        &     (km s$^{-1}$)               &  ($''$)     &  (km s$^{-1}$)       &  (km s$^{-1}$)    &   ($''$)      \\  
  (1)       &     (2)        &   (3)                 &  (4)                & (5)                            &     (6)       &        (7)           &      (8)          &    (9)        \\
\hline
   IC~1815  &  213$\pm$14    &   217                 &  212$\pm$10         &   216                          &  --           &   --                  &  --               &     17.8        \\
   NGC~0043 &  234$\pm$3     &   243                 &  234$\pm$3          &   241                          &  --           &   --                  &  --               &     14.6       \\
   NGC~0098 &  117$\pm$2     &   163                 &  117$\pm$2          &   161                          &  1.5$\pm$0.3 &  134$\pm$1            &  81               &    10.6          \\
   NGC~0175 &  112$\pm$4     &   174                 &  112$\pm$4          &   172                          &  2.6$\pm$0.3 &  126$\pm$1            &  154              &    11.4          \\
   NGC~0521 &  198$\pm$1     &   255                 &  198$\pm$1          &   252                          &  3.5$\pm$0.2  &  201$\pm$2            &  220              &    13.8          \\
   NGC~0621 &  229$\pm$5     &   277                 &  229$\pm$5          &   274                          &  2.7$\pm$0.3  &  234$\pm$2            &  169              &    11.0          \\
   NGC~1640 &  74$\pm$ 7     &   113                 &  74$\pm$2           &   112                          &  1.6$\pm$0.1 &  94$\pm$ 1            &  140              &     18.9         \\
   NGC~2493 &  262$\pm$10    &   284                 &  258$\pm$8          &   281                          &  1.5$\pm$0.4 &  270$\pm$1            &  32               &     27.3         \\
   NGC~4477 &  154$\pm$3     &   180                 &  157$\pm$4          &   179                          &  1.5$\pm$0.1  &  177$\pm$1            &  35               &     45.7          \\
   NGC~4838 &  107$\pm$14    &   145                 &  105$\pm$10         &   144                          &  2.2$\pm$0.1 &  131$\pm$2            &  105              &     18.5          \\
\hline                                 
\end{tabular}
\tablefoot{(1) Galaxy name; (2) and (3) center {\it measured} and {\it
    model corrected}  velocity dispersions, respectively; (4)  and (5)
  {\it  measured} and  {\it  model corrected}  velocity dispersion  at
  $r_{\rm  e}/8$, respectively;  (6)  radius of  the $\sigma-$drop  or
  $\sigma-$plateau; (7) maximum {\it  measured} velocity dispersion of
  the galaxy;  (8) integrated  difference between the  exponential fit
  and  measured  radial  profile   of  the  velocity  dispersion;  (9)
  exponential  scale-length  of the  radial  profile  of the  velocity
  dispersion.}
\end{table*}
  
\begin{figure}[!ht]
   \centering
   \includegraphics[width=0.49\textwidth]{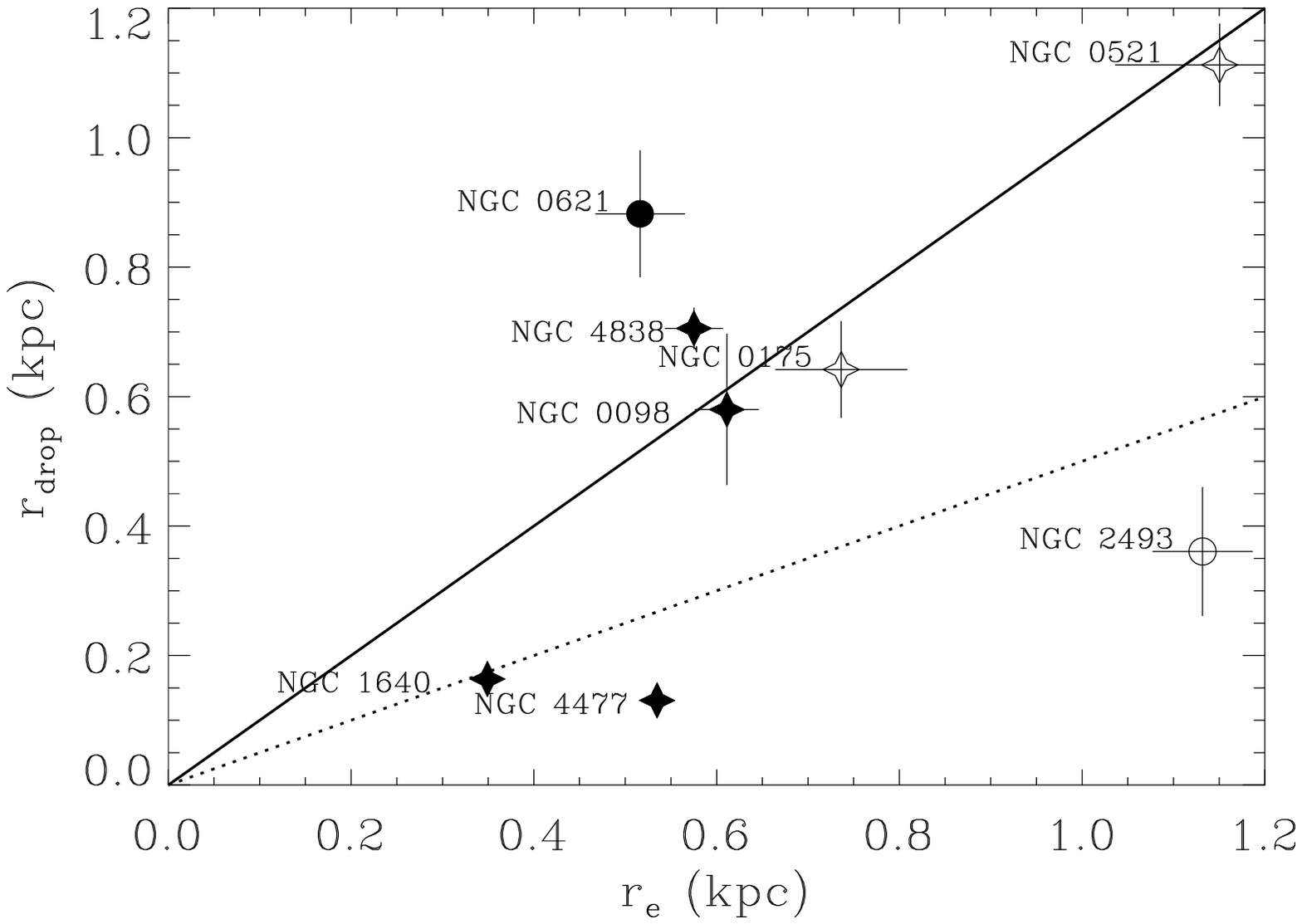}
      \caption{Relation between  the radius  of the  $\sigma-$drop, or
        $\sigma-$plateau, and the effective  radius of the photometric
        bulge.  Filled  and empty  circles represent  classical bulges
        with a drop or a plateau, respectively. Filled and empty stars
        represent pseudobulges with a drop or a plateau, respectively.
        A comprehensive  description of  the bulge  classification for
        each galaxy  is shown in Appendix  \ref{sec:notes}.  The solid
        and  dotted  lines show  both  the  one-to-one and  two-to-one
        relations, respectively.}
         \label{fig:rdropre}
   \end{figure}

Figure \ref{fig:rdropre} shows the relation  between the radius of the
$\sigma-$drop, or  $\sigma-$plateau, and  the effective radius  of the
photometric bulges as defined in Sect.~\ref{sec:decomp}.  As expected,
most of our bulges are larger than the $\sigma-$drops, indicating that
more than  one structural component  is present in the  galaxy center.
However, this  does not hold for  NGC~0621 and NGC~4838.  The  case of
NGC~0621 is  particularly intriguing due  to the double  symmetric dip
present  in  its  velocity  dispersion  profile  (Fig.~\ref{fig:kin}),
pointing out  that the  underlying structure producing  this kinematic
feature is larger than the photometric bulge.  For those galaxies with
$r_{\rm drop}  \sim r_{\rm  e}$ we  suggest that  what is  causing the
$\sigma-$drop can be identified with a pseudobulge.

No correlation was found between the $\sigma-$drop parameters (radius,
peak-to-valley value, or integrated  $\sigma$) and S\'ersic index.  It
is interesting to note, however, that $\sigma-$drops are found only in
galaxies  with $n<2$.   Two out  of  the four  galaxies with  $n\geq2$
present peaked velocity dispersion profiles (IC~1815 and NGC~0043) and
they  are lenticular  galaxies, whereas  the other  two (NGC~0521  and
NGC~2493) show only $\sigma-$plateaus.

\subsection{Central velocity dispersion}
\label{sec:stellar}

We calculated the  central velocity dispersion of  our sample galaxies
in two different ways: first, we directly computed $\sigma_{0}$ as the
measured value of the velocity  dispersion in the center; secondly, we
performed the  luminosity weighted average of  the velocity dispersion
profile   within   an   eighth   of   the   bulge   effective   radius
($\sigma_{r_{\rm e}/8}$) obtained  from the photometric decomposition.
This  quantity is  commonly  used  in the  literature  to compare  the
velocity dispersion of  different systems at the  same physical radius
\citep{jorgensen95}.  These  two values  were also computed  using the
inwards  extrapolation of  the best  exponential fit  to the  velocity
dispersion radial profile.  Thus, we will  refer to the former as {\it
  measured}  and   the  later   as  {\it  model   corrected}  velocity
dispersion, respectively.   In this way,  both values will  be similar
for galaxies  with peaked  velocity dispersion profiles  and different
for $\sigma-$drop, or $\sigma-$plateau  galaxies. It is worth noticing
that due  to the  face-on nature  of the  sample, the  $\sigma$ values
represent  mainly the  vertical component  of the  velocity ellipsoid.
These values are shown in Table \ref{tab:kinproperties}.

\begin{figure}[!ht]
   \centering
   \includegraphics[width=0.49\textwidth]{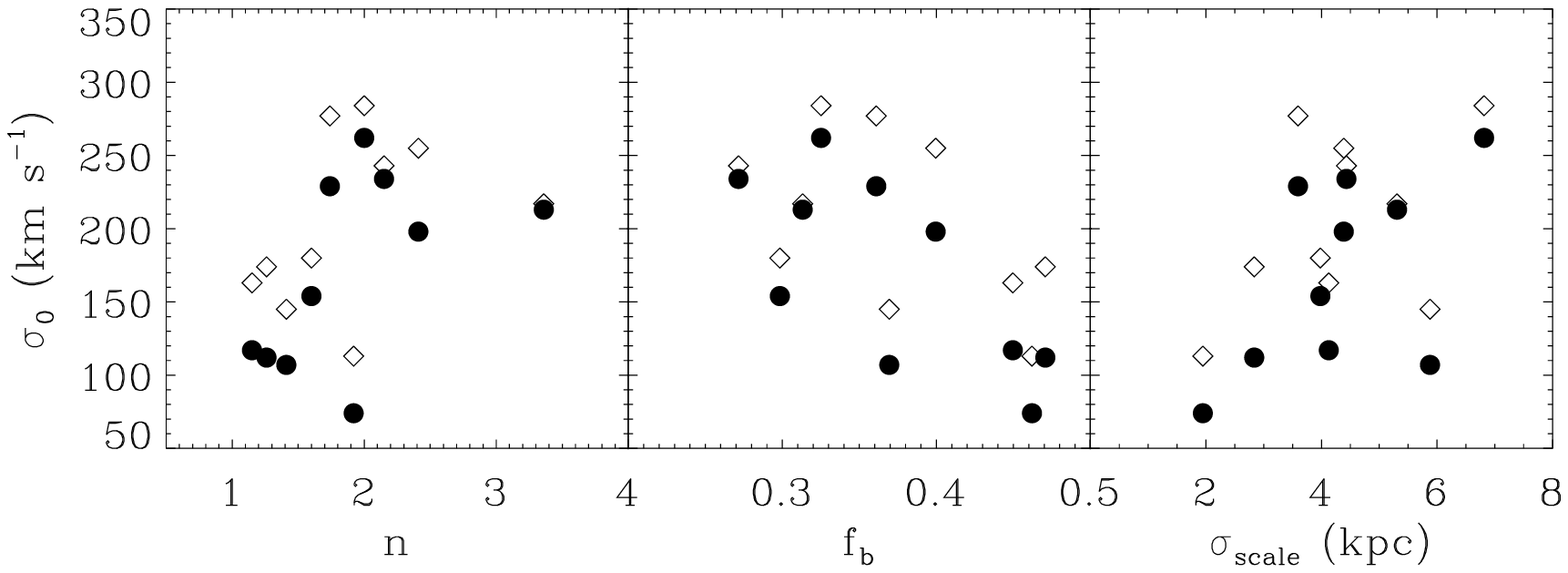}
      \caption{Relation   among   the  central   velocity   dispersion
        $(\sigma_0)$  and  the  S\'ersic  index of  the  bulge  ($n$),
        strength of  the bar  ($f_{\rm b}$), and  scale-length  of the
        velocity  dispersion  radial profile  ($\sigma_{\rm  scale}$).
        Black  dots and  empty diamonds  represent {\it  measured} and
        {\it model corrected} values, respectively.}
         \label{fig:sigma0corr}
   \end{figure}

We  find   a  mild   correlation  (Spearman   correlation  coefficient
$\rho$=0.67) between the  S\'ersic index and the central  value of the
velocity   dispersion  (Fig.    \ref{fig:sigma0corr}).   This   is  in
agreement with  recent results  by \citet{fabricius12} who  found some
differences in the velocity dispersion distribution for bulges with $n
>2.1$ and $n\leq 2.1$. In Fig. \ref{fig:sigma0corr} a clear transition
region around $n\sim2$ and $\sigma_0 \sim 200$ km s$^{-1}$ is visible.

Figure \ref{fig:sigma0corr}  also shows  the correlations  between the
central velocity dispersion  and two bar properties:  the bar strength
($f_{\rm  b}$,  $\rho$=0.75)  and  the scale-length  of  the  velocity
dispersion  radial profile  ($\sigma_{\rm scale}$,  $\rho$=0.63).  The
former  correlation indicates  that strong  bars usually  have smaller
central  velocity  dispersion.  The  latter  shows  how more  radially
extended bar have higher central velocity dispersions.

\subsection{Faber-Jackson relation}
\label{sec:FJ}

\begin{figure}[!ht]
   \centering
   \includegraphics[width=0.49\textwidth]{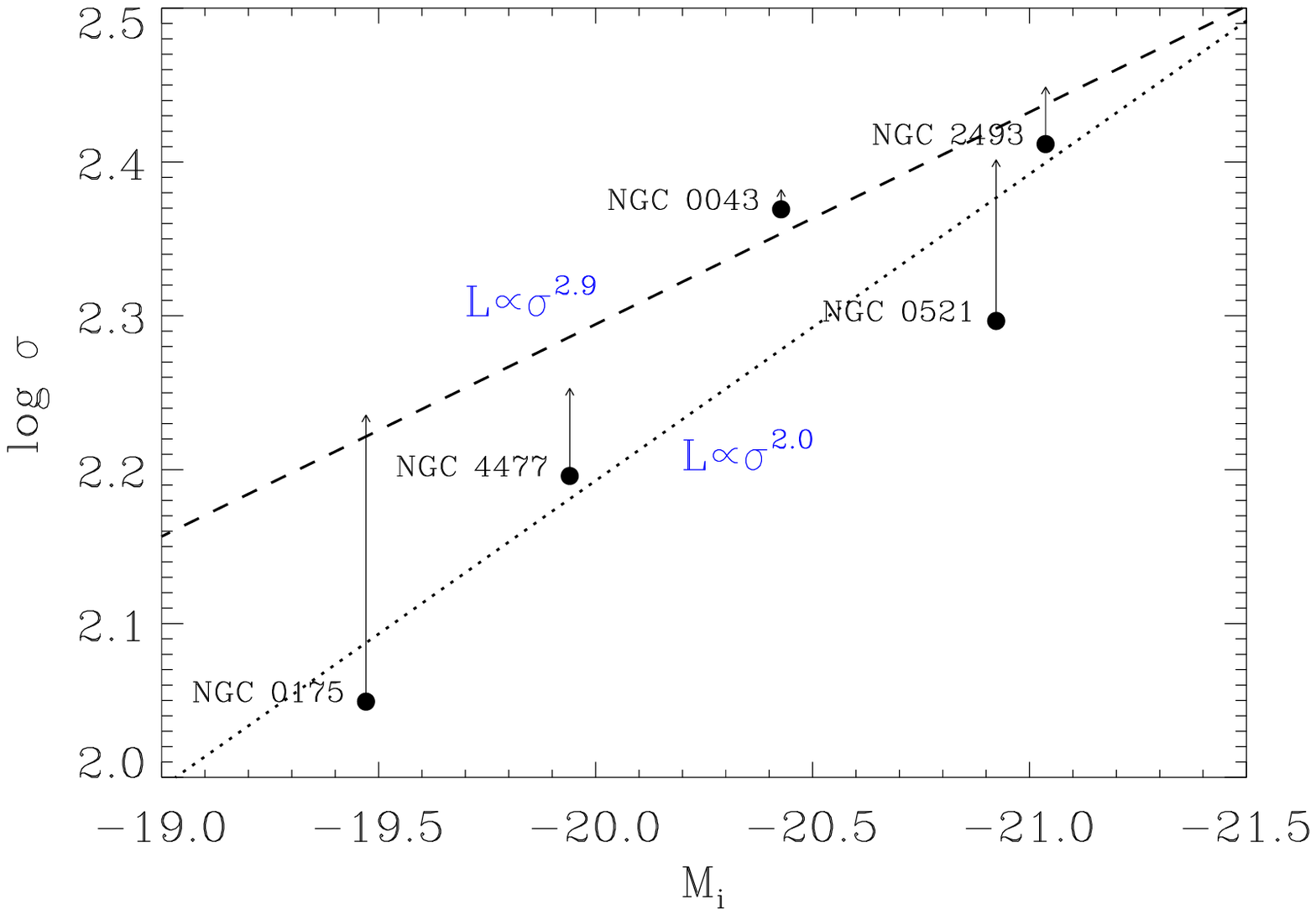}
      \caption{Faber-Jackson relation obtained  for the five bulges of
        the sample with $i-$band SDSS photometry. Black dots represent
        the  {\it measured} values  of $\sigma_{r_{\rm e}/8}$  whereas the
        head  of  the  arrow  marks  the  {\it model corrected}  values  of
        $\sigma_{r_{\rm e}/8}$  obtained from the  exponential fit  to the
        velocity  dispersion radial profile.  Dotted and  dashed lines
        represent the  fits to the {\it measured}  and {\it model corrected}
        FJ relations, respectively.}
         \label{fig:fj}
   \end{figure}

Figure \ref{fig:fj} shows the Faber-Jackson (FJ) relation obtained for
our five bulges  with SDSS photometry in the  $i-$band.  The remaining
galaxies have either uncalibrated photometry  or images in $K-$band so
they are not directly comparable.   The black dots represent the value
of  $\sigma_{r_{\rm  e}/8}$  as   {\it  measured}  from  the  observed
profiles, the head of the arrows show the {\it model corrected} values
of $\sigma_{r_{\rm  e}/8}$ as measured  over the best  exponential fit
performed to  the radial velocity  dispersion profiles. Note  that all
galaxies but  NGC~0043 host  a $\sigma-$drop or  $\sigma-$plateau, and
therefore the differences can be large.

The {\it measured}  FJ relation gives a scaling  $L \propto \sigma^2$,
similar   to   what   is   found   in   low   luminosity   ellipticals
\citep{matkovicguzman05,cody09}.   However,  the  slope for  the  {\it
  model corrected} FJ relation is significantly different with a value
$L \propto  \sigma^{2.9}$, demonstrating the importance  of being able
to resolve the $\sigma-$drops, but  still different from the canonical
value $L  \propto \sigma^{4}$  expected from  the virial  theorem, and
therefore for classical bulges.

Note that  according to Fig. \ref{fig:rdropre},  NGC~0175 and NGC~0521
host  $\sigma-$drops of  dimensions comparable  to their  bulges, thus
they   can   be   associated   to   disk-like   structures   such   as
pseudobulges.  In  Fig.~\ref{fig:fj},  they  also fall  below  the  FJ
relation,     as     expected     if     they     are     pseudobulges
\citep{kormendykennicutt04}.

\subsection{B/P bulges through spectroscopy and morphology}
\label{sec:bp}

B/P  bulges are  usually detected  in  edge-on galaxies  due to  their
characteristic  shape.    The  pioneering   surveys  carried   out  by
\citet{desouza87} and \citet{shaw87} demonstrated that they are common
structures observed in edge-on  galaxies.  More recently, the presence
of these structures have been unambiguously related to the presence of
a bar component \citep{bureaufreeman99,lutticke00}.  However, recently
different methodologies have been used to identify these structures in
less  inclined galaxies.   \citet{debattista05}  proposed a  kinematic
diagnostic to  search for B/P  bulges in  face-on galaxies based  on a
double minimum in  the $h_4$ radial profile along the  bar major axis.
\citet{mendezabreu08b}  observationally confirmed  this prediction  on
NGC~0098 where a  clear double minimum in $h_4$ is  present before the
end of the bar. We search  for similar features in the radial profiles
of our sample  of face-on galaxies (Fig.  \ref{fig:kin}).   We did not
find any other  example as clear-cut as NGC~0098 in  the whole sample.
Some  galaxies  show marginal  detections  such  as NGC~1640  ($r_{\rm
  B/P}\simeq8\arcsec$),  NGC~2493  ($r_{\rm B/P}\simeq10\arcsec$),  or
NGC~0043  ($r_{\rm  B/P}\simeq5\arcsec$,  but   visible  only  on  the
approaching side).   Two possibilities  can be  argued to  explain the
non-detections: first,  not all barred  galaxies host a B/P  bulge and
therefore  the non-detection  of minima  could  imply that  no B/P  is
present in those galaxies; second, it  is possible that the quality of
our  galaxy  spectra is  not  sufficient  to disentangle  more  subtle
features  in  the  $h_4$   profiles  (see  \citealt{debattista05}  for
examples of the $h_4$ minima variations).

\begin{figure*}[!ht]
   \centering
   \includegraphics[width=\textwidth]{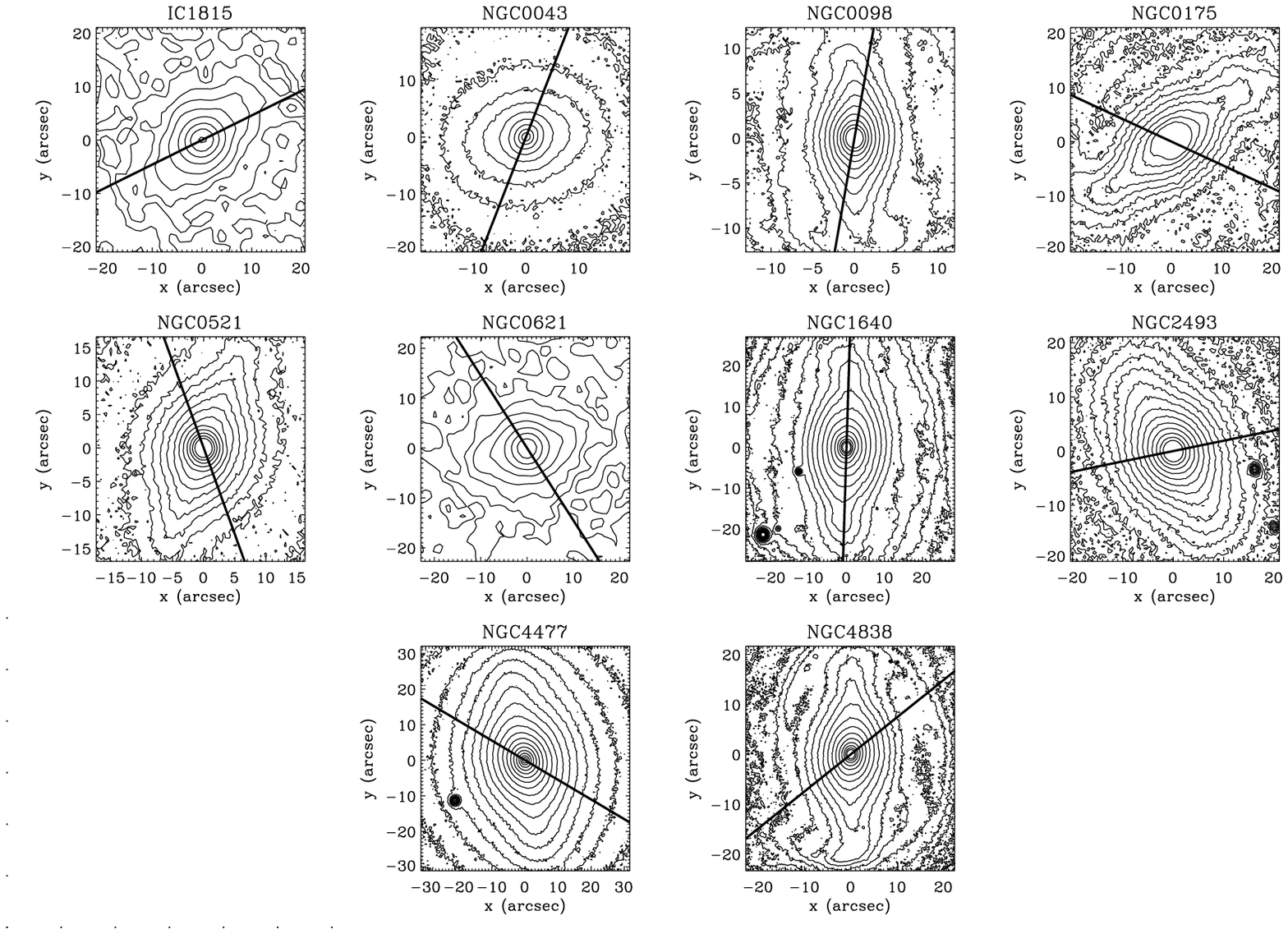}
      \caption{Logarithmically   scaled   isophotes  of   the   sample
        galaxies.  Each  panel covers  roughly the  bar region  of the
        galaxy and  the solid  line represents the  major axis  of the
        galaxy disk.}
         \label{fig:iso}
   \end{figure*}

Recently,  \citet{erwindebattista13} demonstrated  the possibility  of
detecting B/P  structures in  moderately inclined galaxies  using just
their  morphology.   Their  method  is   based  on  the  detection  of
characteristic  features  in  the  galaxy  isophotes  created  by  the
projection  of the  thick B/P  structure combined  with a  thinner and
larger bar.  In  practice, the bar isophotes are made  of two regions:
the interior of the bar is broad and slightly boxy in shape, while the
outer part  of the bar forms  narrower spurs, almost always  offset or
even rotated with respect to the major axis of the inner, boxy region.
Even if  our galaxies are,  by choice, face-on rather  than moderately
inclined, we analyzed  the isophotes of the galaxies in  our sample in
order to  search for these  features.  In Fig.  \ref{fig:iso}  we show
the isophotes for our sample galaxies.  The only marginal case of this
morphology is  found in NGC~0175,  which has inner boxy  isophotes and
offset  spurs.  The  radius  of  this inner  boxy  structure is  about
7$\arcsec$, which roughly corresponds to  a minimum in the approaching
side  of  the  $h_4$  radial   profile.   However,  we  did  not  find
photometric signs of a B/P  bulge in NGC~0098.  Therefore, we conclude
that at these inclinations the  photometric detection of B/P bulges is
strongly      dependent      on     the      bar-disk      orientation
\citep{erwindebattista13}.

\section{Discussion}
\label{sec:discussion}

\subsection{A zoo of bulge types}

As  commented previously,  bulges  come in  different  flavours and  a
careful  case-by-case   study  is   needed  to  unveil   their  real
nature. However, the  most commonly used recipe to  classify bulges in
their  different  types  is   by  analyzing  their  surface-brightness
distribution.   Since pseudobulges  share  many  properties of  galaxy
disks,  it  is  expected  that their  surface  brightness  follows  an
exponential profile.   On the other  hand, classical bulges  which are
expected to better resemble the properties of ellipticals might follow
a more  steep profile.   In terms of  the S\'ersic  parametrization of
bulges, pseudobulges should  have a S\'ersic index  $n\sim1$ whereas a
classical  bulge  should  have  $n\sim4$.  This  diagnostic  has  been
proposed   as    a   good    proxy   for   separating    bulge   types
\citep{kormendykennicutt04,
  droryfisher07,fisherdrory08,fisherdrory11}.                Recently,
\citet{fabricius12} carried  out an extensive study  of how structural
properties  such as  the S\'ersic  index  are related  to the  stellar
kinematics. They  found that both  properties are correlated  and that
photometric  properties can  be  used to  safely  determine the  bulge
nature.   However,  morpho-kinematics  relations  show  a  significant
degree of  scatter and a safe  classification of bulge type  should be
done on an individual basis.

We   performed   an  accurate   case-by-case   study   based  on   the
morphological,  photometric, and  kinematic properties  of bulges.   A
crucial  point in  our analysis  is that  the LOS  velocity dispersion
probed in this work can be  directly related to the vertical component
of the velocity ellipsoid.  Since  pseudobulges are associated to disk
structures,  and  these  are  supposed to  be  dynamically  cold,  the
vertical component of the velocity ellipsoid in pseudobulges should be
low enough  to maintain a disk  structure. We apply this  condition to
our classical vs. pseudobulge selection by imposing that all classical
bulges should have $\sigma_0 > 200$ km\,s$^{-1}$.  This limit is based
on our  relation between the  S\'ersic index and the  central velocity
dispersion  of bulges  (Fig.  \ref{fig:sigma0corr})  which shows  that
around  $n\sim2$  and  $\sigma_0  >   200$  km\,s$^{-1}$  there  is  a
transition region  that might  be used  to distinguish  classical from
pseudobulges in disk  galaxies.  In fact, the S\'ersic  index was also
used  in  our  classification  by  imposing  a  boundary  line  around
$n\sim2$. However, a larger sample  of galaxies with vertical velocity
dispersions measured is needed to confirm this result.

One possible  source of  scatter in  previous works  using statistical
limits  to separate  bulge types  might  be the  presence of  multiple
structures  in  the galaxy  centers.   We  carefully checked  for  the
presence of composite bulges in  our sample.  Particular attention was
paid  to  the  presence  of  B/P  structures  as  explained  in  Sect.
\ref{sec:bp}. We found hints, either photometric or kinematic, for B/P
bulges  in 5  galaxies of  the sample  (NGC~0043, NGC~0098,  NGC~0175,
NGC~1640, and NGC~2493).   In all these cases the  estimated radius of
the B/P  components was  much larger that  the bulge  effective radius
indicating   that   they   co-exist    with   another   classical   or
pseudobulge. However, it is worth noting that the detection of the B/P
structures is marginal in some cases.  On the other hand, the presence
of $\sigma-$drops, or $\sigma-$plateaus, is  also used in this work to
identify embedded pseudobulges.  Fig.  \ref{fig:rdropre} shows how the
extension of the $\sigma-$drops can  be comparable, larger, or smaller
than  the photometric  bulge effective  radius therefore  implying the
presence of a  single, or composite bulge in our  sample galaxies.  In
particular,  when the  effective radius  of the  photometric bulge  is
larger than the $\sigma-$drops,  or $\sigma-$plateaus, we assume there
is a  nuclear cold  component (usually  considered as  a pseudobulge).
Then, the properties of the region attributed to the photometric bulge
define  the   classification  of  the  secondary   bulge.   Particular
attention was paid to assure  the presence of two different components
in the  center of these  galaxies (NGC~1640, NGC~2493,  and NGC~4477),
and discard a  possible fake detection of the nuclear  disk due to the
exponential  fading of  the surface  brightness.  We  exclude possible
light  contamination effects  caused by  other galaxy  components. The
photometric bulge  effective radius is  between two to four  times the
size   of   the    central   $\sigma-$drops,   or   $\sigma-$plateaus.
Furthermore, the $r_{\rm bd}$, i.e.,  the radius where the bulge light
dominates over every  other galaxy component \citep[see][]{morelli08},
spans from  three to  nine times  the size  of the  $\sigma-$drops and
$\sigma-$plateaus.   Finally, we  compute the  $B/T$ luminosity  ratio
within one $r_{\rm e}$ finding that  at least 75\% of the galaxy light
at  this  radius comes  from  the  photometric bulge.   Therefore,  we
suggest that the nuclear disk and  the outer photometric bulge are not
the same structure.

\begin{figure}[!ht]
   \centering
   \includegraphics[width=0.49\textwidth]{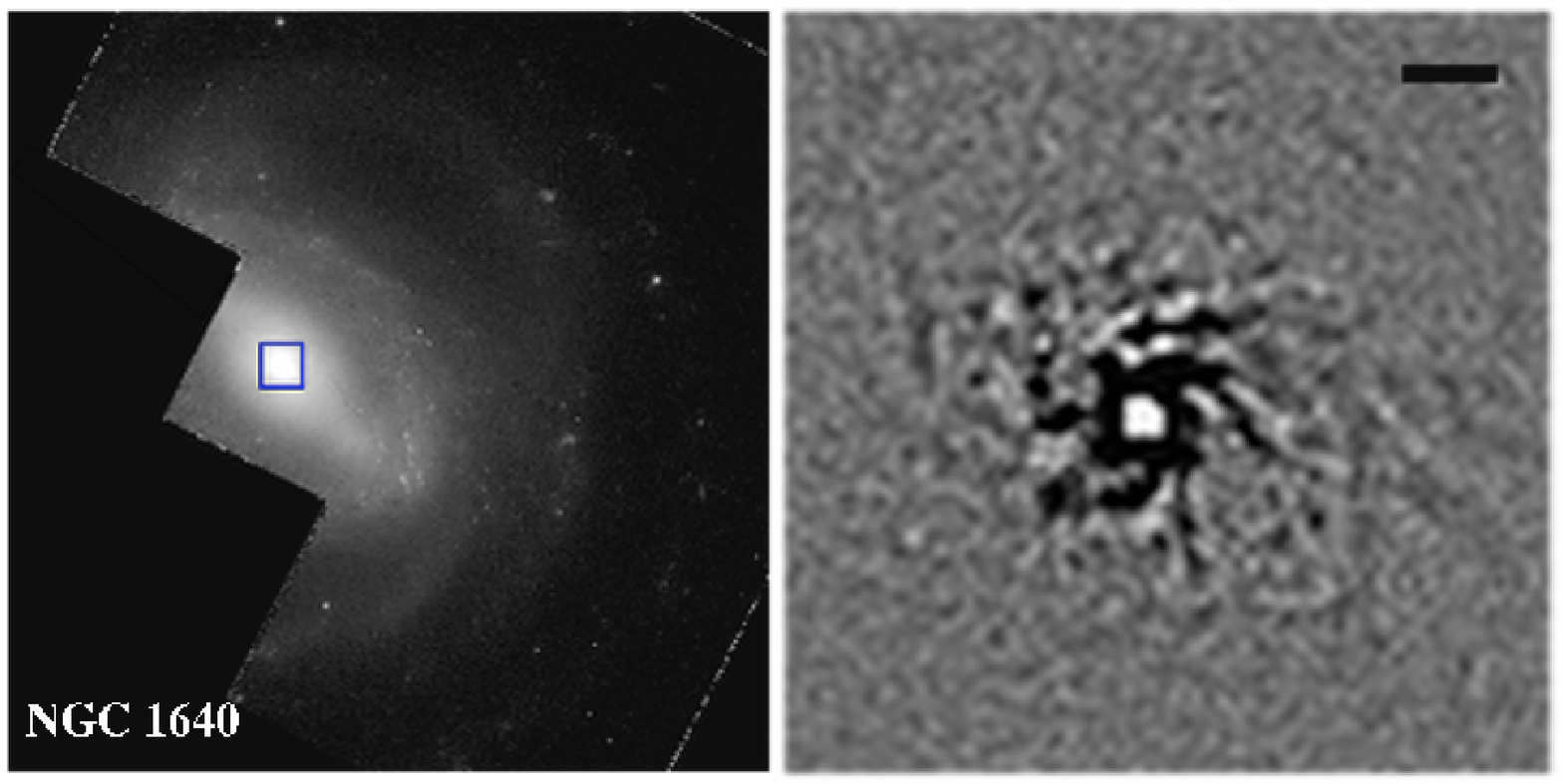}
   \includegraphics[width=0.49\textwidth]{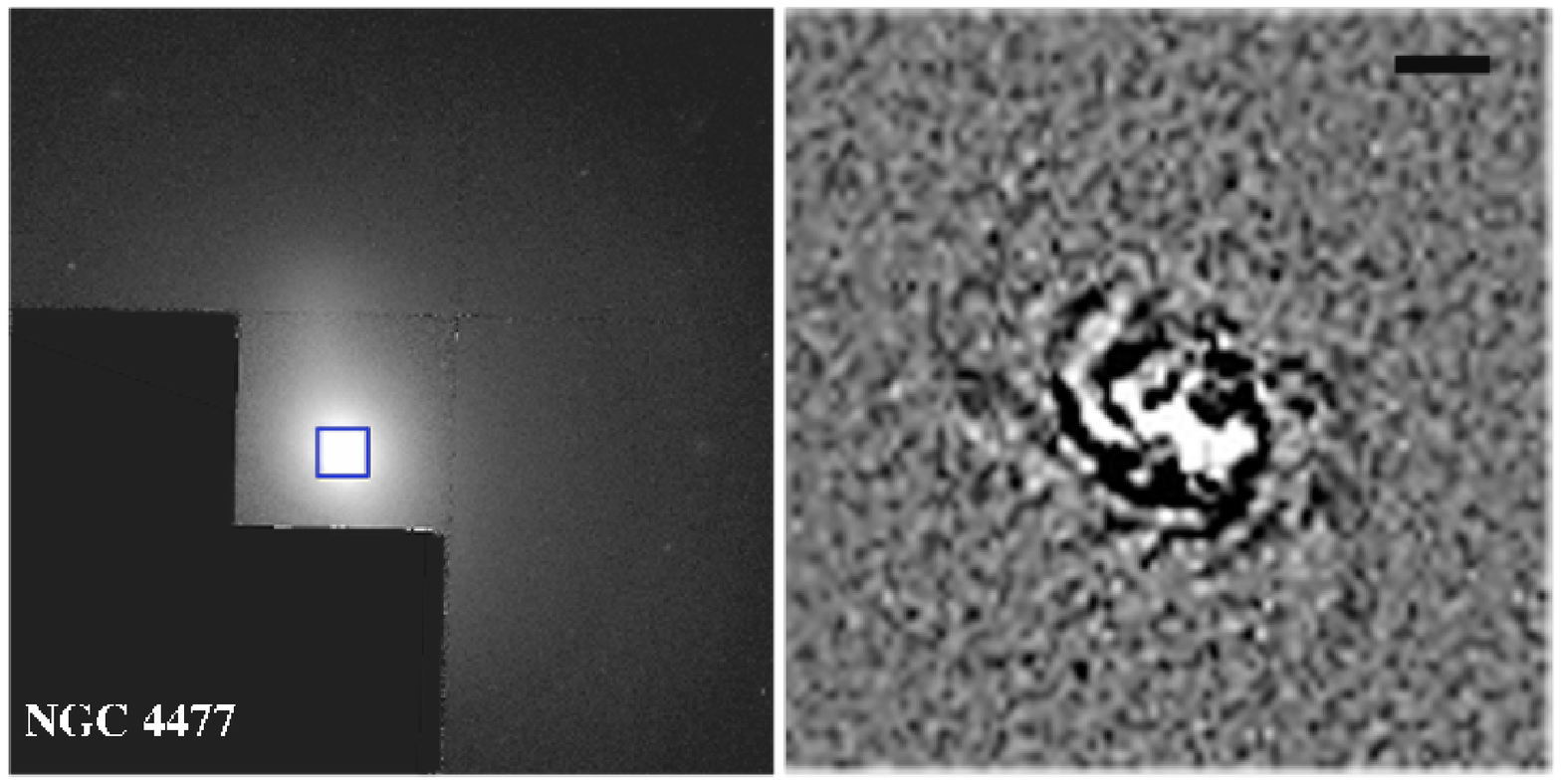}
      \caption{Mosaic  including the  {\it HST}  image in  the $F606W$
        band (left  panels) and  a zoomed unsharp  image of  the inner
        regions (right panels). Mosaics  are shown for NGC~1640 (upper
        panels) and NGC~4477 (lower panels).   The blue squares in the
        galaxy   centers  (left   panels)  represent   the  size   and
        orientation of  the unsharp  masked image (right  panels). The
        scale bars at the top represent 1\arcsec.}
         \label{fig:unsharp}
   \end{figure}

High  spatial   resolution  images  represent  a   powerful  tool  for
investigating the presence  of different structures in  the very inner
regions of  disk galaxies.   They have  been used  in some  studies to
reveal the presence of nuclear spirals, nuclear bars, and nuclear star
forming rings  which are indicative  of the presence of  a pseudobulge
\citep{kormendykennicutt04}.   Unfortunately, in  our sample  only two
galaxies have  been observed with  {\it HST} (NGC~1640  and NGC~4477).
Both of  them host  a central  cold component  clearly visible  in the
velocity  dispersion profile  as a  $\sigma-$drop with  a spiral  disk
counterpart   showing    up   in   the   {\it    HST}   images   (Fig.
\ref{fig:unsharp}).   However, the  photometrically defined  bulges of
these  galaxies  are  much  larger  than  the  nuclear  spiral  disks,
independently confirming that those bulges  are, at least, composed of
two different  structures.  In  the two  galaxies with  available {\it
  HST} images, we suggest the  photometric bulge is a pseudobulge, but
this  should be  distinguished from  the nuclear  disk.  On  the other
hand, some galaxies (NGC~0098, NGC~0175,  and NGC~0521) have a central
cold component, identified from their $\sigma-$drop, which corresponds
in     size     with     the     photometrically     defined     bulge
(Fig.~\ref{fig:rdropre}).  In this  case the direct link  of a central
disk  with pseudobulges  seems  more obvious,  even  though {\it  HST}
images would be necessary to confirm this structural counterpart.

Appendix A  gives a complete  description of the bulge  properties and
classification  obtained for  each galaxy  of the  sample.  The  final
classification   as   well   as  the   main   bulge   morpho-kinematic
characteristics are summarized in Table \ref{tab:bulgetype}.

\begin{table*}
\caption{Bulge types hosted by our sample galaxies}   
\label{tab:bulgetype}    
\centering  
\resizebox{0.99\textwidth}{!}{
\begin{tabular}{c c c c c c c c c c c c}      
\hline\hline
Galaxy   &                       &  CLASSICAL  &                          & &                      &  PSEUDO&                          &      &\hspace{1cm}B/P  &       & Comments \\
\cline{2-4}
\cline{6-8}
\cline{10-11}
         & $\sigma>200$ km s$^{-1}$          & $n\gtrsim2$ & $r_{\rm drop} < r_{\rm e}$ & &$\sigma<200$ km s$^{-1}$          & $n<2$ & $r_{\rm drop} \gtrsim r_{\rm e}$ & &$h_4$& Spurs &          \\
  (1)    &      (2)              &       (3)   &           (4)            & &    (5)               &   (6) &         (7)               &      & (8) &  (9)  &  (10)    \\
\hline
IC~1815  &      y                &     y        &            n/a          & &           n          &    n  &        n/a                &      &  n  &  n     & Classical   \\
NGC~0043 &      y                &     y        &            n/a          & &           n          &    n  &        n/a                &      &  y  &   n    & Classical + B/P?\\
NGC~0098 &      n                &     n        &            n            & &           y          &    y  &         y                 &      &  y  &  n     & Pseudo + B/P\\
NGC~0175 &      n                &     n        &            n            & &           y          &    y  &         y                 &      &  y  &  y     & Pseudo + B/P\\
NGC~0521 &      n                &     y        &            n            & &           y          &    n  &         y                 &      &  n  &  n     & Pseudo\\
NGC~0621 &      y                &     n        &            n            & &           n          &    y  &         y                 &      &  n  &  n     & Classical + inner bar\\
NGC~1640 &      n                &     n        &            y            & &           y          &    y  &         n                 &      &  y  &  y     & Central spiral disk + pseudo + B/P?\\
NGC~2493 &      y                &     y        &            y            & &           n          &    n  &         n                 &      &  y  &   n    & Central cold structure + classical + B/P?\\
NGC~4477 &      n                &     n        &            y            & &           y          &    y  &         n                 &      &  n  &   n    & Central spiral disk + pseudo\\
NGC~4838 &      n                &     n        &            n            & &           y          &    y  &         y                 &      &  n  &  n     & Pseudo\\
\hline                                 
\end{tabular}
}
\begin{minipage}{0.95\textwidth}
\tablefoot{(1) Galaxy name;  from (2) to (7) is  the galaxy fulfilling
  this condition? yes  or no; (8) Does the galaxy  show h$_4$ minima?
  yes or no; (9) Does the  galaxy show photometric spurs? yes or no;
  (10) Composition of  the galaxy bulges and other  structures based on
  their morpho-kinematic characteristics.}
\end{minipage}
\end{table*}

\subsubsection{Single-component bulges}

Only  3 out  of 10  galaxies  in our  sample have  been identified  as
hosting a  single-component bulge  (IC~1815, NGC~0521,  and NGC~4838).
This implies  that 70\% of  our galaxies host composite  bulges.  This
result proves the  complexity of bulge structures  in barred galaxies.
Moreover, the fraction of single-component bulges should be considered
as a upper  limit since fainter nuclear structures not  resulting in a
$\sigma$-drop,  or  $\sigma-$plateau,  can  still be  present  in  the
galaxies.   Two galaxies,  NGC~0521 and  NGC~4838, host  a pseudobulge
indicating  a  possible secular  formation,  most  likely due  to  the
presence  of a  bar \citep{kormendykennicutt04}.   On the  other hand,
IC~1815 hosts a  classical bulge.  A likely scenario  for the assembly
of  classical  bulges implies  their  formation  through major  merger
events.  Therefore,  this galaxy  hints at an  early formation  of the
bulge  before  the  bar  forms  in a  rebuilt  disk  after  the  major
merger. Another possible scenario could  indicate a joint formation of
both  components  due to  un-equal  mass  mergers or  interactions  as
proposed by \citet{perez09} for early-type barred galaxies.

\subsubsection{Composite secular-built bulges}

We  refer to  {\it  composite secular-built  bulges}  as those  bulges
composed by  two or more  structures associated to  secular processes,
namely pseudobulges, central  disks, or B/P bulges.   Four galaxies in
our sample  host a composite secular-built  bulge (NGC~0098, NGC~0175,
NGC~1640, and  NGC~4477).  Taking into account  the previously noticed
single-component pseudobulges,  60\% of  our sample consist  of bulges
formed mainly by secular processes.

Three galaxies (NGC~0098, NGC~0175, and NGC~1640) show features linked
to the presence of a B/P structure in addition to the pseudobulge. B/P
structures are  associated to  the evolution of  bars in  galaxy disks
\citep{combessanders81,martinezvalpuesta06}, and  therefore it  is not
surprising to find  the co-existence of both components  in the center
of barred  galaxies.  However, due to  inclination-related problems on
detecting  B/P   components  in   relatively  face-on   galaxies,  the
identification    of    these     composite    systems    have    been
elusive. \citet{kormendybarentine10} detected one  of these systems in
the  edge-on   galaxy  NGC~4565  using  near-infrared   imaging.  They
demonstrated that the  B/P structure is associated to  a bar structure
with  a tiny  pseudobulge contributing  only  about 6\%  to the  total
galaxy    light.      A    similar    study    was     presented    by
\citet{barentinekormendy12} for NGC~5746. The present work opens a new
way to study  these composite systems in face-on  galaxies, allowing a
more  straightforward detection  of different  bulge types  within the
same galaxy.

Other secular-built  central structures, namely central  spiral disks,
have been also  identified in two galaxies (NGC~1640  and NGC~4477) of
our  sample using  available  {\it HST}  observations.  These  central
spiral disks  are also  clearly visible  in their  velocity dispersion
profiles as a  $\sigma-$drop and their dimensions  are consistent with
being  the same  structure.  On  the other  hand, the  photometrically
defined bulges are much larger than the nuclear spiral disks and their
properties are  consistent with being pseudobulges.   Thus, we suggest
these galaxies formed their pseudobulge  and also their central spiral
disk through secular processes.

\subsubsection{Composite merger- and secular-built bulges}

We consider  {\it merger-  and secular-built  bulges} as  those bulges
composed of  a merger-built  bulge, namely a  classical bulge,  and at
least one secular-built structure, i.e., pseudobulge, central disk, or
B/P.   Three  galaxies (NGC~0043,  NGC~0621,  and  NGC~2493) from  our
sample satisfy  these criteria showing evidences  for composite bulges
associated to different formation  mechanisms.  These bulges represent
a challenge to formation models  that need to accommodate merger-built
classical  bulges with  either central  rotationally supported  or bar
associated structures.

We have found the co-existence of a classical and B/P structure in two
galaxies, NGC~0043  and NGC~2493.   Different formation  scenarios for
these bulges  include an early  bulge formation through  major mergers
with  a subsequent  bar  development on  a rebuilt  disk  or a  coeval
formation  of bulge  and bar  triggered  by un-equal  mass mergers  or
interactions  \citep{perez09}.   The  presence of  the  B/P  structure
imposes a further  constraint on the timeline of  bulge evolution. The
usual time for a bar to be settled in a galaxy disk can vary depending
on  the  galaxy  and  halo properties  \citep{athanassoula13}  but  it
generally takes about 1-2 Gyr.  Taking into account that the formation
of B/P structures also imply a delay of about 1-2 Gyr after the bar is
formed  \citep{martinezvalpuesta06}. This  would  imply our  classical
bulges in  these two lenticular galaxies  are at least older  than 3-4
Gyr.  In  addition, this  result  can  be  interpreted as  bars  being
long-lived structures  as claimed in  previous papers based  on theory
\citep{kraljic12},    kinematics   \citep{gadotti05},    and   stellar
populations \citep{gadotti06,sanchezblazquez11}.

NGC~0621   presents  a   double,  symmetric   drop  of   the  velocity
dispersion. These drops are located farther than the photometric bulge
and therefore  cannot be  associated with  an embedded  component. Our
most likely  explanation of  the drops  is due to  the presence  of an
inner bar. Thus, the drops of the velocity dispersion are connected to
the sigma  hollows discovered by  \citet{delorenzocaceres08}. However,
other possibilities  such as the presence  of an inner ring  cannot be
ruled  out, but  its  position  outside the  bulge  region makes  this
scenario less likely.

Our galaxies  show composite bulges structurally  different from those
described  in  \citet{nowak10}.   They found  small  classical  bulges
embedded  in   larger  pseudobulges,  with  this   happening  even  in
double-bar galaxies (i.e., NGC~3368). However, we always found a large
photometrically defined  classical bulge with an  inner cold structure
related  to a  pseudobulge. This  description is  more similar  to the
scenario proposed  by \citet{delorenzocaceres12} for the  bulge of the
double-barred galaxy NGC~0357.

\subsection{Origin of $\sigma-$drops}
We  found $\sigma-$drops  in five  galaxies of  our sample:  NGC~0098,
NGC~0621, NGC~1640, NGC~4477, and NGC~4838.  Central $\sigma-$plateaus
were  found   in  three   other  galaxies:  NGC~0175,   NGC~0521,  and
NGC~2493. This means that 80\% of our sample galaxies do not present a
centrally peaked velocity dispersion.  The incidence of $\sigma-$drops
in  barred   galaxies  is  still   an  open  debate.    For  instance,
\citet{perez09}  found a  70\% of  galaxies hosting  $\sigma-$drops in
their sample  of barred  galaxies whereas  \citet{chungbureau04} found
only 40\%.  In mixed, barred  and non-barred, samples the incidence of
$\sigma-$drops  is   around  50\%  \citep{falconbarroso06,peletier07}.
Therefore, the  high fraction  of $\sigma-$drops  found in  our sample
seems  to  indicate  a  connection between  bars  and  $\sigma-$drops,
contrary to the previous claims  by \citet{comeron08}.  However, it is
not  clear whether  our sample  is  too small  to extract  statistical
conclusions  or  if  the  medium-resolution  spectroscopy  on  face-on
galaxies more easily reveals these structures.

The standard explanation  for these $\sigma-$drops is  the presence of
central    cold    stellar    disks    formed    from    gas    inflow
\citep{emsellem01,wozniak03}.     \citet{wozniakchampavert06}    using
dynamical simulations, predicted that these features can last for more
than 1  Gyr if  the central  region is continuously  fed by  fresh gas
leading to  a continuous  star formation  activity.  On  the contrary,
\citet{athanassoulamisiriotis02}   found   $\sigma-$drops  in   N-body
dissipationless  simulations, and  \citet{bureauathanassoula05} showed
that  they can  arise from  the orbital  structure of  strongly barred
galaxies. Another possibility is  that the central $\sigma$-drop might
be  caused  by  the  bulge itself,  which  therefore  shows  disk-like
properties that indicate  it is a pseudobulge rather  than a classical
one   \citep{delorenzocaceres12}.   Within   this  new   context,  the
$\sigma$-drop  is not  actually  a  drop with  respect  to the  higher
velocity dispersion of the bulge,  but the maximum velocity dispersion
of the pseudobulge itself.

Only two  galaxies in our  sample have high spatial  resolution images
from the {\it HST}.  Both of  them host a $\sigma-$drop: NGC~1640 and
NGC~4477 (Fig.~\ref{fig:unsharp}).  After  unsharp masking the images,
a  nuclear disk  with spiral  arms showed  up in  both galaxies.   The
dimensions of these structures are in  good agreement with the size of
the $\sigma-$drop.   This seems to  favour the scenario in  which cold
nuclear  stellar disks  can originate  from gas  inflow along  the bar
\citep{emsellem01, wozniak03}.  It is  worth noting that these nuclear
disks are not directly connected with the photometric bulge, therefore
indicating the presence of two coexistent components.

On the other hand, three  galaxies show $\sigma-$drops with dimensions
comparable  to  that of  their  host  bulges.  Therefore,  these  cold
structures can be associated with the  presence of a pseudobulge as in
the scenario proposed by \citet{delorenzocaceres12} for NGC~0357.

Finally,  \citet{fabricius12}  suggest   that  classical  bulges  show
preferentially peaked velocity  dispersion profiles while pseudobulges
have $\sigma-$plateaus or even $\sigma-$drops.  We confirm this result
with our  face-on sample.  Two out  of the four classical  bulges show
peaked  velocity  dispersion  profiles  in  our  sample  (IC~1815  and
NGC~0043).  NGC~2493  also hosts  a classical  bulge and  presents the
shallower plateau of the sample, being recognizable only after fitting
an  exponential model  to the  velocity dispersion  profile.  NGC~0621
shows a symmetrical double $\sigma-$drop  which we interpret as due to
the presence of a nuclear bar.

\section{Conclusions}
\label{sec:conclusion}

We have  studied the  photometric and  kinematic properties  of bulges
using  a   sample  of  10   face-on  barred  galaxies.    An  accurate
two-dimensional photometric  decomposition was  applied to  the galaxy
images in  order to derive  the structural properties of  their bulges
and bars.   Intermediate resolution  long-slit spectroscopy  along the
bar major axis was used to derive the LOSVD radial profiles.

We  found  a rough  correlation  among  the structural  and  kinematic
properties  for  different bulge  types,  similar  to the  results  by
\citet{fabricius12}.   Even if  the  scatter is  high, we  tentatively
found that a S\'ersic index  $n\sim2$ combined with a central vertical
velocity dispersion  $\sigma_0 \sim 200$ km\,s$^{-1}$  provides a good
boundary line between classical and pseudobulges.

A careful case-by-case analysis of  the bulges allowed us to identify,
and  classify,  the  different  bulge  types  present  in  our  sample
galaxies:  classical  vs  pseudobulges. In  addition,  B/P  structures
associated with the  presence of a bar were  also identified.  Special
attention was  paid to the  presence of composite bulges.   Indeed, we
distinguish between  single-component bulges  and composite  bulges in
our sample.  We found only 3  single-component bulges in our sample: 2
are  pseudobulges and  1  is a  classical bulge.   The  presence of  a
classical bulge in  a barred galaxy implies that it  was likely formed
before the bar.

We found  evidence for the presence  of multiple bulges within  the so
called  ``bulge region''.   These composite  bulges have  already been
found         in         a         handful         of         galaxies
\citep{erwin03,nowak10,delorenzocaceres12}.     However,   the    high
incidence (70\%)  of these  composite bulges  in our  sample indicates
that mixed types might be very  frequent in barred galaxies.  We found
four secular-built  composite bulges  with structures  compatible with
being formed by  secular processes. More interestingly,  we also found
three  bulges where  merger- and  secular-built bulges  co-exist. This
kind  of composite  bulges  is not  completely  surprising, and  their
presence  within  barred  galaxies  have already  been  considered  in
theoretical studies \citep{samlandgerhard03,athanassoula05}.  However,
the merger- and  secular-built bulges found in our sample  is new, and
different from  previous works  \citep{nowak10}, since we  found large
classical bulges  with smaller pseudobulges coexisting  in the nuclear
regions.

We found a high fraction (70\%) of $\sigma-$drops or $\sigma-$plateaus
in our  sample. These  $\sigma-$drops are  usually connected  with the
presence  of  a  rapidly  rotating  structure in  the  center  of  the
galaxies.  We confirm  this interpretation  in two  galaxies with  HST
imaging where a clear spiral disk, comparable in size with that of the
$\sigma-$drops, is found.  In some cases the size of the $\sigma-$drop
is comparable with  the size of the photometric bulge  while in others
the bulge can be  more than a factor of two  larger.  This might imply
different origins for the $\sigma-$drops and the presence of different
coexisting structures within the same bulge, respectively.

\begin{acknowledgements}
We thank P.  Erwin and  A. de Lorenzo-C\'aceres for useful discussions
and  suggestions.  We  thank  the  referee  for  his/her  constructive
comments which helped to improve the  paper. This work has been partly
funded   by   the   Spanish   Ministry  for   Science,   project   AYA
2010-21887-C04-04.   JMA   acknowledges  support  from   the  European
Research Council  Starting Grant (SEDmorph;  P.I.  V.  Wild).   VPD is
supported  by  STFC  Consolidated  grant  \#ST/J001341/1.  E.   M.  C.
acknowledges  financial support  from Padua  University by  the grants
60A02-5052/11,  60A02-4807/12, and  60A02-5857/13.  Funding  for SDSS,
SDSS-II,  and SDSS-III  has  been  provided by  the  Alfred P.   Sloan
Foundation,  the  Participating  Institutions,  the  National  Science
Foundation, and the U.S.  Department of Energy Office of Science. This
publication makes  use of data  products from  the Two Micron  All Sky
Survey, which  is a joint  project of the University  of Massachusetts
and the  Infrared Processing and Analysis  Center/California Institute
of  Technology,   funded  by   the  National  Aeronautics   and  Space
Administration and the National  Science Foundation. This research has
made  use  of the  NASA/IPAC  Extragalactic  Database (NED)  which  is
operated  by the  Jet Propulsion  Laboratory, California  Institute of
Technology,  under contract  with the  National Aeronautics  and Space
Administration.
\end{acknowledgements}
\bibliographystyle{aa} 
\bibliography{reference} 

\begin{appendix}
\section{Notes on individual galaxies}
\label{sec:notes}

\noindent
{\bf IC~1815:}  Only 2MASS imaging  is available for this  galaxy. The
S\'ersic index  of the bulge ($n=3.4$)  is the highest of  our sample.
The   velocity   profile  shows   a   mild   double-hump  profile   at
$r=3.5\arcsec$ but the velocity dispersion  does not show any evidence
of  flattening or  depression  in the  central  regions.  The  central
velocity  dispersion is  larger than  200 km  s$^{-1}$ suggesting  the
bulge  is pressure  supported.  We  suggest this  galaxy hosts  a {\it
  classical bulge}.

\noindent
{\bf  NGC~0043:}  The SDSS  image  shows  a smooth  surface-brightness
distribution with  the weakest bar  of our sample. The  bulge S\'ersic
index is $n=2.2$.   No clear signs of a double-hump  rotation curve or
depression  in  the  central  velocity dispersion  are  found.   These
characteristics  point  towards the  presence  of  a classical  bulge.
However,  the $h_4$  radial  profile is  asymmetric,  showing a  clear
minimum   on  the   approaching  side   of  the   galaxy  at   $r_{\rm
  B/P}=6\arcsec$.  This corresponds to $\sim$1/3 of the bar length, in
agreement with  the findings of \citet{mendezabreu08b}  and suggesting
the presence  of a  B/P bulge.   We suggest this  galaxy hosts  a {\it
  classical bulge} and possibly a {\it B/P bulge} too.

\noindent
{\bf  NGC~0098:} The  residuals of  the photometric  decomposition are
large due  to the prominence  of the spiral  arms. The best  fit bulge
model has an effective radius $r_{\rm e}=1.6\arcsec$ which corresponds
fairly well with the size of  the $\sigma-$drop present in the stellar
kinematics ($r_{\rm drop} = 1.5\arcsec$).   The rotation curve shows a
step  rise in  this inner  region and  anti-correlates with  the $h_3$
moment.  The S\'ersic  index ($n=1.2$) is compatible  with a disk-like
structure.  The spectroscopic observations of this galaxy were already
analyzed in  \citet{mendezabreu08b} and the characteristic  minimum in
the $h_4$ radial profile produced by a B/P bulges was found at $r_{\rm
  B/P}=5\arcsec$.  We suggest this galaxy hosts a {\it pseudobulge} at
the very center and {\it B/P bulge} farther out.

\noindent
{\bf NGC~0175:}  Also known as NGC~0171  in NED.  A truncated  disk is
clearly  visible in  its SDSS  surface-brightness profile.   The bulge
accounts for  only 7\% of  the total  galaxy light, in  good agreement
with the result obtained by \citet{buta09}.  They used deep $K_s-$band
imaging to  study the bar  properties and suggested that  an elongated
inner ring is  present in the galaxy center.   The velocity dispersion
profile shows a central plateau which does not correspond to any clear
increasing  rotation.    \citet{ho07}  reported  a   central  velocity
dispersion  $\sigma_0$=104 km\,s$^{-1}$  which agrees  with our  value
($\sigma_0$=112  km\,s$^{-1}$).  The  size  of the  plateau is  barely
smaller ($r_{\rm  drop}=2.6\arcsec$) than  the bulge  effective radius
($r_{\rm  e}=3.0\arcsec$).  In  Sect.~\ref{sec:bp}  we speculate  also
about the presence of B/P structure at $r_{\rm B/P}=7\arcsec$ based on
its morphological and kinematic features.  We suggest that this galaxy
hosts  an  {\it  inner  ring}  that   can  be  associated  to  a  {\it
  pseudobulge} and an outer {\it B/P bulge}.

\noindent
{\bf  NGC~0521:} A  truncated  disk  is clearly  visible  in the  SDSS
surface-brightness profile.   \citet{buta09} found the presence  of an
inner ring with a radius $r=4.3\arcsec$ \citep[see also][]{comeron10}.
The velocity  dispersion radial profile  shows a central  plateau with
$r_{\rm drop}=3.5\arcsec$,  which corresponds  to a  steep rise  of the
rotation curve  (showing a  marked double-hump profile).  The velocity
anti-correlates  with $h_3$.   The bulge  effective radius  is $r_{\rm
  e}=3.6\arcsec$.   We suggest  this galaxy  hosts a  central rotating
structure that can be associated with a {\it pseudobulge}.

\noindent
{\bf NGC~0621:} Only 2MASS imaging  was available for this galaxy. The
velocity  dispersion radial  profile in  the center  shows a  peculiar
double  drop,  symmetric  with  respect to  the  galaxy  center.  This
behavior of the velocity dispersion is similar to the $\sigma-$hollows
discovered  in \citet{delorenzocaceres08},  which are  related to  the
presence of an  inner bar.  These $\sigma-$hollows have  turned out to
be more common than expected and  have been detected in other galaxies
\citep{delorenzocaceres12}.  Moreover,  the size of  the $\sigma-$drop
($r_{\rm drop}=2.7\arcsec$) also point  towards the inner bar scenario
since  it is  much larger  that  the bulge  effective radius  ($r_{\rm
  e}=1.6\arcsec$).   We  consider  that   the  high  central  velocity
dispersion  obtained  for  this  galaxy, which  suggests  there  is  a
classical bulge,  fits well in the  scenario of a double  bar, since a
high velocity dispersion contrast is  a necessary condition to observe
the  $\sigma-$hollows  \citep{delorenzocaceres08}.   Nevertheless,  we
cannot rule out  other possibilities such as the presence  of an inner
ring.  Therefore, we suggest this galaxy  hosts an {\it inner bar} (or
{\it inner ring}) and possibly a {\it classical bulge}.

\noindent
{\bf NGC~1640:} The SDSS image shows the presence of strong dust lanes
along the bar major axis.  This is confirmed by the $F606W-$band image
of the {\it  HST}/WFPC2.  In fact, this high  spatial resolution image
shows how  the dust lanes  converge to  the galaxy center  where inner
spiral  arms  can  be  clearly   seen  in  the  unsharp  masked  image
(Fig.~\ref{fig:unsharp}).   The radius  of  this  inner structure,  as
measured from the WFPC2 image, is $r=1.5\arcsec$, which corresponds to
the  size of  the $\sigma-$drop  ($r_{\rm drop}=$1.6\arcsec).   On the
contrary,  the   bulge  effective  radius  is   much  larger  ($r_{\rm
  e}=$3.4\arcsec).  The $h_4$  radial profiles show hints  of a double
minimum at $r_{\rm B/P}\simeq8\arcsec$, i.e.,  at $\sim$1/6 of the bar
radius.   We suggest  that this  galaxy  hosts a  {\it central  spiral
  disk}, a  slightly larger {\it  pseudobulge}, and possibly  an outer
{\it B/P bulge}.

\noindent
{\bf NGC~2493:} No sign of dust obscuration is seen in the SDSS image.
There  is a  mild plateau  in the  center of  the velocity  dispersion
profile which can be better distinguished when comparing with the best
fit exponential model extrapolation.  The size of the plateau ($r_{\rm
  drop}=1.5\arcsec$) is  much smaller than the  bulge effective radius
($r_{\rm e}=$4.6\arcsec).  The  $h_4$ radial profile shows  hints of a
double minimum at $r_{\rm  B/P}\simeq10\arcsec$ (i.e., at $\sim$1/3 of
the bar  radius).  We suggest  that this  galaxy hosts a  {\it central
  cold structure}, a  {\it classical bulge} and possibly  a {\it B/P
  bulge}.
 
\noindent 
{\bf   NGC~4477:}  The   residuals  obtained   from  the   photometric
decomposition  of  this galaxy  are  quite  low, indicating  that  the
parametric model for the bulge, bar,  and disk describes the SBD well.
The central  region of  this galaxy  is unveiled thanks  to the  { \it
  HST}/WFPC2 image  in the $F606W$  band.  Strong dust  obscuration is
seen throughout the bar region and  inner spiral arms are clearly seen
in the  unsharp masked image (Fig.~\ref{fig:unsharp}).   The radius of
this  structure  is  $r=1.5\arcsec$  which corresponds  to  the  small
$\sigma-$drop  ($r_{\rm  drop}=1.5\arcsec$)  present in  the  velocity
dispersion profile.  We suggest that this galaxy hosts an {\it central
  spiral disk} and a {\it pseudobulge}.

\noindent
{\bf NGC~4838:} The velocity dispersion  radial profile shows a strong
$\sigma-$drop in the center. The rotation curve is strongly asymmetric
and  presents also  a step  rise but  only on  the receding  side, the
rotation   curve  being   strongly  asymmetric.    The  size   of  the
$\sigma_{\rm  drop}$ ($r_{\rm  drop}=$2.2\arcsec) roughly  corresponds
with the effective  radius of the bulge  ($r_{\rm e}=$1.8\arcsec).  We
suggest that this galaxy hosts a {\it pseudobulge}.
\end{appendix}

\end{document}